\documentclass[preprint,12pt]{elsarticle}

\usepackage{caption}
\usepackage{amsmath}
\usepackage{amssymb}
\usepackage{amsthm}
\usepackage{mathrsfs}

\usepackage{epstopdf}

\usepackage{subfig}

\usepackage{booktabs} 
\usepackage{multirow} 

\usepackage{url}
\usepackage{hyperref}
\usepackage{xcolor}
\usepackage{cleveref} 

\usepackage{lineno}

\usepackage{threeparttable}
\usepackage{color}
\usepackage{epstopdf}
\usepackage{float}
\usepackage[top=2.5cm, bottom=2.5cm, left=2cm, right=2cm]{geometry}
\renewcommand{\vec}[1]{\boldsymbol{#1}}

\biboptions{numbers,sort&compress}

\journal{}

\begin{document}


\begin{frontmatter}



\title{Radiative hydrodynamic equations with nonequilibrium radiative transfer}

\author[a]{Mingyu Quan}
\author[a]{Xiaojian Yang}
\author[a]{Yufeng Wei}
\author[a,b,c]{Kun Xu\corref{cor1}}

\cortext[cor1] {Corresponding author.}
\ead{makxu@ust.hk}

\address[a]{Department of Mathematics, Hong Kong University of Science and Technology, Hong Kong, China}
\address[b]{Department of Mechanical and Aerospace Engineering, Hong Kong University of Science and Technology, Hong Kong, China}
\address[c]{HKUST Shenzhen Research Institute, Shenzhen, 518057, China}

\begin{abstract}

 This paper presents a kinetic model for the coupled evolution of radiation, electrons, and ions in a radiation plasma system. 
 The model is solved using two methods. The gas-kinetic scheme (GKS) for electron and ion hydrodynamics and the unified gas-kinetic scheme (UGKS) 
 for non-equilibrium radiative transfer. The UGKS accurately captures multiscale photon transport from free streaming to diffusion across varying fluid opacities. 
 This approach enables the scheme to model equilibrium plasma with non-equilibrium radiation transport.
 The model is validated through several test cases, including radiative transfer in kinetic and diffusion regimes, Marshak wave, Radiative shock, 3T(three-temperature) double lax shock tube problem, two-dimensional Sedov blast wave, and two-dimensional tophat based problem.
These tests demonstrate the current scheme's capability to handle diverse radiation plasma scenarios.
 
\end{abstract}


\end{frontmatter}



\section{Introduction}\label{sec:intro}
Interactions and energy exchange processes among electrons, ions, and radiation have significant implications in various fields, including astrophysics, weapon physics, and inertial confinement fusion (ICF) \cite{drake2006,mihalas1999}. In astrophysics, stellar interiors host high-temperature plasmas where electrons and ions undergo intense collisions in the continuum regime. The radiation field generated through nuclear fusion in stars is influenced by convective motion, magnetic field movement, and matter absorption/scattering, leading to a highly non-equilibrium state. This non-equilibrium radiation plays a crucial role in the cooling of electrons and ions, thereby impacting the energy generation mechanisms and evolutionary processes of stars \cite{di1997two,jiang2019super}. In weapon physics, the detonation of high-energy density materials generates high-temperature plasmas that undergo rapid energy release under the influence of strong magnetic and electric field effects. This leads to significant variations in the relaxation frequencies between electrons and ions. In the presence of a strong radiation intensity background, this high-energy non-equilibrium relaxation process will affect the energy release, hydrodynamic/thermodynamic characteristics, and the interactions with the surrounding environment of the explosive material \cite{atzeni2004}. Inertial Confinement Fusion (ICF) utilizes high-energy lasers or particle beams to achieve nuclear fusion reactions. In this process, radiation decouples from electron and ion temperatures, resulting in anisotropic radiation intensity and non-equilibrium spectral, electrons and ions are also in a highly non-equilibrium state \cite{Fraley1974}. The non-equilibrium spectra of radiation, together with the non-equilibrium states of electrons and ions, constitute a multiscale non-equilibrium transport across different regimes, significantly impacting fuel compression and energy transport processes. Therefore, the development of multiscale numerical simulation methods capable of describing the interactions among electrons, ions, and radiation is of crucial importance for understanding complex non-equilibrium physical processes in multi-physics.

Numerical simulation methods for radiation plasma systems composed of electrons, ions, and photons can be divided into two approaches: microscopic stochastic particle methods and numerical solutions of macroscopic hydrodynamic equations.
When the applications require transport-type modeling, one of the most used methods for thermal radiation propagation is the Implicit Monte-Carlo (IMC) method \cite{wollaber2016four}. The ion-electron coupling and conduction can be split or coupled fully or partially during the Monte Carlo calculation of the radiation-transport process \cite{evans2007methods}. However, a linear stability analysis shows that the coupling of the IMC method with non-equilibrium ion/electron physics introduces larger, damped temporal oscillations, and nearly unity amplification factors can lead to a divergent temperature \cite{Wollaber2013}. Additionally, the time step satisfied a discrete maximum principle should be adopted \cite{wollaber2013discrete} since the IMC method can produce nonphysical overshoots of material temperatures when large time steps are used.

To balance the efficiency and accuracy, macroscopic radiation hydrodynamic is adopted when the applications under consideration allow diffusion-type approximations.
Therefore, the evolution of the interaction between electrons, ions, and radiation field can be described by the three-temperature (3-T) radiation hydrodynamics (RH) equations, which is a reduced model for radiative transport coupled with plasma physics. The 3-T RH equations consist of the advection, diffusion, and energy-exchange terms which are highly nonlinear and tightly coupled. Numerical methods on the three-temperature radiation hydrodynamic have become an active research topic in recent years. A 3-T RH numerical method based on the multi-group strategy is developed by reformulating and decoupling the frequency-dependent 3-T plasma model \cite{Enaux2022}. The method is suited for handling strict source terms while significantly reducing the numerical cost of the method and preserving basic discrete properties. An extremum-preserving finite volume scheme for the 3-T radiation diffusion equation \cite{Peng2022}. The scheme has a fixed stencil and satisfies local conservation and discrete extremum principles. The high-order scheme for 3-T RH equations has also been developed. A third-order conservative Lagrangian scheme in both spatial and temporal space is designed based on the one-dimensional 3-T RH equations \cite{cheng2024}. The scheme possesses both conservative properties and arbitrary high-order accuracy and will be primarily applied in situations where the timescale is much larger than the photon transport.

In radiation hydrodynamics, the process involves the propagation of radiation through a moving material. The movement of the material introduces a velocity component, which necessitates the inclusion of relativistic corrections in the thermal radiative transfer equation. These corrections are particularly important when the deposition of radiation momentum significantly affects the dynamics of the material. The correction is applicable for flow with moving velocity being much smaller than the speed of light.
Lowrie\cite{lowrie1999} solves the RT
equation in the mixed frame: the specific intensity is measured in an Eulerian frame while the radiation–material interaction terms are computed in the co-moving frame of the fluid. This requires
a Lorentz transformation of variables between frames. This provides a suitable set of equations for nonrelativistic radiation hydrodynamics
(RHD) that can be numerically integrated using high-resolution methods for conservation laws.  Holst \cite{van2011crash} propose a block-adaptive-mesh code for multi-material radiation hydrodynamics where a single fluid description is used so that
all of the atomic and ionic species, as well as the electrons, move with the same bulk velocity $u$. This 3-D code can be applied to astrophysics and laboratory astrophysics. Sun \cite{sun2020multiscale} proposed a system that couples the fluid dynamic equations with the radiative heat transfer.

The coupled system is solved by the gas-kinetic scheme (GKS) for the compressible inviscid Euler flow and the unified gas-kinetic scheme (UGKS) for the non-equilibrium radiative transfer, together with the momentum and energy exchange between these two phases. This system is capable of naturally capturing the transport process from the photon’s free streaming to the diffusive wave propagation. A new algorithm for solving the coupled frequency-integrated transfer equation and the equations of magnetohydrodynamics in the regime that light-crossing time is only marginally shorter than dynamical timescales is described by Jiang \cite{jiang2014global}. This couples the solution of the radiative transfer equation to the MHD algorithms in the Athena code and has an important influence on simulating the black hole accretion disks\cite{jiang2014radiation,jiang2019global}.

For the governing equation depicting flow physics, different from molecular dynamics and hydrodynamic equations, the kinetic model can effectively describe both the microscopic molecular motion processes and the macroscopic flow states. Therefore, in this paper, the kinetic model is used to describe the radiation plasma system. The model proposed treats electrons and ions as a binary fluid system represented by the multispecies model. Radiation is considered a background field that mediates energy exchange with electrons. The relaxation processes in this model are manifested as transport and diffusion. The hydrodynamics limit of the method can yield the same equations as the traditional radiation hydrodynamic equations.

For the numerical simulation methods, different from microscopic stochastic particle methods and numerical solution of macroscopic hydrodynamic equations, in this paper, the unified gas-kinetic scheme (UGKS) is used to solve the kinetic model of the radiative transfer equation \cite{xu2010,xu2014}. The UGKS was first applied to solve neutral gas transport processes. Its main idea is based on a finite volume framework that directly models the flow processes at numerical spatiotemporal scales. The flux evolution functions are constructed by integrating gas kinetic model equations. By simultaneously coupling the free transport and collision processes in the evolution, the method overcomes the limitations of grid size and time step imposed by the average mean free path and mean collision time in splitting methods such as the Direct Simulation Monte Carlo (DSMC) \cite{bird1994molecular} method and the Discrete Velocity Model (DVM) method \cite{broadwell1964study}. The UGKS exhibits favorable asymptotic properties and, based on the mesoscopic BGK model, can achieve efficient solutions in the entire flow regime. It yields results consistent with the Navier-Stokes (NS) equations in the continuum flow regime and with the DSMC method in the free-molecular flow regime. UGKS has been extended to solve complex physical fields involving real gas effects \cite{liu2014unified,wang2017unified} and multispecies \cite{xiao2019unified,wang2014}. In particular, UGKS has been successfully applied to solve plasmas composed of ions and electrons \cite{liu2017unified}. Additionally, the transport-diffusion processes caused by radiation transfer in both optically thin and optically thick are also multiscale problems. The radiation transfer equation bears a striking resemblance to the gas kinetic BGK model in its formulation. Therefore, the UGKS approach can be directly extended to the numerical solution of radiation transport equations as well \cite{mieussens2013}. In the construction of interface flux evolution functions, the integration solution of the radiation transport equation is used to couple radiation transport and radiation diffusion processes. It has also been extended to grey radiation, multi-frequency radiation, and other complex radiation transport problems \cite{sun2015gray,sun2015frequency}.

In this paper, the radiation plasma system is firstly introduce and asymptotic properties are analyzed in Section \ref{sec:model}. Then the numerical simulation schemes of the model using GKS-UGKS will be presented in Section \ref{sec:scheme}. Finally, the proposed models and their numerical methods will be thoroughly validated through comprehensive simulations of the shock tube problem, tradiative transfer problem, radiative shock problem, and two-dimensional Sedov problem in Section \ref{sec:case}. In Section \ref{sec:conclusion}, the conclusion of the paper is given.

\section{The kinetic model for radiation plasma system}\label{sec:model}
The kinetic model for the radiation plasma system and the asymptotic analysis will be discussed in this section. The kinetic model for multispecies gas mixture is adopted for ion and electron, and the radiative transfer equation considering the momentum and energy exchange with the ion and election two-fluid system is introduced to solve the radiation process in the model. The interactions within the radiation plasma system in the model and its asymptotic analysis of the hydrodynamic limits will also be introduced in detail.

\subsection{Kinetic model}

A gas kinetic theory-based kinetic model is employed to describe the radiation plasma system. The electrons and ions will be treated as two distinct fluids, utilizing the single-relaxation gas kinetic model proposed by Andries et al. \cite{aap2002}. This kinetic model for multispecies mixture satisfies the indifferentiability principle, and entropy condition, and can recover the exchanging relationship of Maxwell molecules with such a simple relaxation form. When the radiation momentum deposit has a measurable impact on the material dynamics, the thermal radiative transfer equation needs to be corrected based on the material velocity. The modification is needed even for the case where the speed of flow is much smaller than the speed of light. Under such a condition, the $M M(\theta)$ model \cite{lowrie1999} for the coupled radiation and hydrodynamics is adopted in the current study. For non relativistic flows, keeping only terms to $\mathcal{O}(\vec{U}/c)$ \cite{mihalas1999}. The kinetic model can be expressed as
\begin{equation} \label{eq:kinetic-model}
\begin{aligned}
	&\frac{\partial f_\alpha}{\partial t} + {\vec u}\cdot\frac{\partial f_\alpha}{\partial {\vec x}} = \frac{g_{\alpha}^M - f_\alpha}{\tau_\alpha}, \qquad \alpha \in \{ \mathcal{E}, \mathcal{I}\}, \\
	&\frac{\epsilon}{c}\frac{\partial I}{\partial t} +\vec{\Omega} \cdot \frac{\partial I}{\partial \vec{x}} + \epsilon\frac{\partial }{\partial \vec{x}}\cdot (\theta \vec{\beta} I)= \frac{\sigma_t}{\epsilon}\biggl(\frac{1}{4\pi}acT^{4}-I\biggr) + \frac{\epsilon \sigma_{s}c}{4\pi}(E_{\mathcal{R}} - aT^{4})  \triangleq S_{\mathcal{R}},
\end{aligned}
\end{equation}
where $f_{\alpha} = f (\vec{x},\vec{u},t)$ is the distribution function for molecules at physical space location $\vec{x}$ with microscopic translational velocity $\vec{u}$ at time $t$. The mean collision time or relaxation time $\tau_{\alpha}$ represents the mean time interval of two successive collisions. The subscription $\alpha$ belongs to a set of species with electron $\mathcal{E}$, ion $\mathcal{I}$, and $\mathcal{R}$ represents radiation. $I(\vec{x},\vec{\Omega},t)$ is the radiation intensity, spatial variable is denoted by $\vec{x}$, $\vec{\Omega}$ is the angular variable, and $t$ is the time variable. For simplicity, in this paper, we only consider the gray case, where the intensity is integrated over the radiation frequency. $\sigma_{t}$ is the total coefficient of absorption, $\sigma_{s}$ is the coefficient of scattering, $a$ is the radiation constant, and $c$ is the speed of light, $\epsilon > 0$ is the optical thickness. $T$ is the electron temperature, $\vec{U}$ is the fluid velocity, $\vec{\beta} \equiv {\vec{U}}/{c}$ and $\theta$ is a free variable related to the correction due to the material motion, $\theta = 4/3$ is used in this paper which can be viewed as an approximation comoving-frame treatment\cite{sun2020multiscale}.
The equilibrium state $g_{\alpha}^{M}$ gives
\begin{equation}\label{eq:eq-state}
g_{\alpha}^{M}=\rho_{\alpha}\left(\frac{m_{\alpha}}{2\pi k_BT_{\alpha}^{M}}\right)^{3/2}\exp\left(-\frac{m_{\alpha}}{2k_{B}T_{\alpha}^{M}}\left(\vec{u}-\vec{U}_{\alpha}^{M}\right)^{2}\right),
\end{equation}
where $k_B$ is the Boltzmann constant, and $m_\alpha$ is the molecular mass of the species.
The macroscopic velocity $\vec{U}_\alpha^M$ and temperature $T_\alpha^M$ is given after momentum and energy exchange by multispecies collisions between election and ion, denoted by the superscription $\star$
\begin{equation}\label{aap-exchange}
\begin{aligned}
	&\vec{U}_{\alpha}^\star = \vec{U}_{\alpha} + \tau_{\alpha}\sum\limits^N\limits_{k = \mathcal{E}}\frac{2\rho_{k}\chi_{\alpha k}}{m_{\alpha} + m_k}(\vec{U}_{k} - \vec{U}_{\alpha}), \\
	&T_{\alpha}^\star = T_{\alpha} - \frac{m_{\alpha}}{3k_{B}}(\vec{U}_{\alpha}^\star - \vec{U}_{\alpha})^2  + \tau_{\alpha}\sum\limits^N\limits_{k = \mathcal{E}}\frac{4\rho_{k}\chi_{\alpha k}}{(m_{\alpha} + m_k)^2}\left(T_{k} - T_{\alpha} + \frac{m_{k}}{3k_{B}}(\vec{U}_{k} - \vec{U}_{\alpha})^2\right),
\end{aligned}
\end{equation}
and the momentum and energy exchange with radiative, denoted by the superscription $M$
\begin{equation}
\begin{aligned}
	& \vec{U}_{\mathcal{E}}^M=\vec{U}_{\mathcal{E}}^{\star} - \frac{\tau_{\mathcal{E}}}{c\rho_{\mathcal{E}}} \int \vec{\Omega} S_{\mathcal{R}} {\rm d} \vec{\Omega}, \\
	& \vec{U}_{\mathcal{I}}^M=\vec{U}_{\mathcal{I}}^{\star} , \\
	& T_{\mathcal{E}}^M=T_{\mathcal{E}}^{\star}-\frac{m_{\mathcal{E}}}{3 k_B}\left[\left(\vec{U}_{\mathcal{E}}^M\right)^2-\left(\vec{U}_{\mathcal{E}}^{\star}\right)^2\right]- \frac{2m_{\mathcal{E}}\tau_{\mathcal{E}}}{3\epsilon \rho_{\mathcal{E}} k_{B}} \int S_{\mathcal{R}} {\rm d} \vec{\Omega}, \\
	& T_{\mathcal{I}}^M=T_{\mathcal{I}}^*, \\
\end{aligned}
\end{equation}
where the set species $N = \left\{\mathcal{I}, \mathcal{E}\right\}$ and the macroscopic velocity and temperature before the inelastic collisions
\begin{equation}\label{multispecies property2}
\begin{aligned}
	&\vec{U}_{\alpha} = \frac{1}{\rho_{\alpha}} \int \vec{u}f_\alpha {\rm d} \vec{u},\\
	&T_{\alpha}=\frac{m_\alpha}{3\rho_{\alpha}k_{B}}\int \left(\vec{u}-\vec{U}_{\alpha}\right)^{2}f_\alpha {\rm d} \vec{u}, \quad \alpha \in \mathcal{I}, \mathcal{E}.
\end{aligned}
\end{equation}
The relaxation time $\tau_\alpha$ is calculated by
\begin{equation}\label{eq:aap-tau}
\frac{1}{\tau_{\alpha}}=\sum_{k=\mathcal{E}}^{N}\chi_{\alpha k}n_{k},
\end{equation}
with the number density $n_\alpha$ of the species and the multispecies interaction coefficient given by hard sphere model
\begin{equation}\label{eq:aap-chi}
\chi_{\alpha k}=
\frac{4\sqrt{\pi}}{3}\Bigl(\frac{2k_{B}T_{\alpha}}{m_{\alpha}}+\frac{2k_{B}T_{k}}{m_{k}}\Bigr)^{1/2}\Bigl(\frac{d_{\alpha}+d_{k}}{2}\Bigr)^{2},
\end{equation}
where $d_\alpha$ is the molecular diameter of the species $\alpha$.
In this model, the relaxation process between ion and radiation is ignored.

\subsection{Asymptotic analysis}
The second equation of the model\eqref{eq:kinetic-model} has the property to approach the equilibrium diffusion limit equations when $\epsilon \to 0$ under any choice of $\theta$ \cite{sun2020multiscale}.
First expanding the variables as a power series of $\epsilon$,
\begin{equation}\label{rad-expand}
\vec{U} = \sum^{\infty}_{i = 0} \vec{U}^{i}\epsilon^{i},
\quad
T = \sum^{\infty}_{i = 0} T^{i}\epsilon^{i},
\quad
I = \sum^{\infty}_{i = 0} I^{i}\epsilon^{i},
\end{equation}
and substituting Eq.\eqref{rad-expand} into the Eq.\eqref{eq:kinetic-model}, then collecting the terms with same powers. The $\mathcal{O}(\epsilon^{-1})$ terms give
\begin{equation}
I^0 = \frac{1}{4\pi}ac(T^0)^4,
\end{equation}
and then $E_{\mathcal{R}}^{0} = a(T^{0})^4, \vec{F}_{\mathcal{R}}^{0} = 0$ and $\vec{P}_{\mathcal{R}}^{0} = \frac{1}{3}a(T^{0})^4 \vec{\rm{D}}$ are obtained by
\begin{equation}
E_{\mathcal{R}} = \frac{1}{c}\int I {\rm d} \vec{\Omega},
\quad
\vec{F}_{\mathcal{R}} = \int \vec{\Omega} I {\rm d} \vec{\Omega},
\quad
\vec{P}_{\mathcal{R}} = \frac{1}{c}\int \vec{\Omega} \times \vec{\Omega} I {\rm d} \vec{\Omega},
\end{equation}
where $\vec{\rm D}$ is the identity matrix. The $\mathcal{O}(\epsilon^0)$ terms give
\begin{equation}
I^{1} = \frac{1}{4\pi}ac(T^1)^4 - \frac{c}{\epsilon_{t}}\vec{\Omega}\cdot\partial_{\vec{x}} I^{0} ,
\end{equation}
followed by
\begin{equation}\label{rad-ap1st}
E_{\mathcal{R}}^1 = a(T^{1})^4,
\quad
\vec{F}_{\mathcal{R}}^1 = -\frac{c}{3\sigma_{t}}\vec{\Omega}\cdot\partial_{\vec{x}} E_{\mathcal{R}}^{0},
\quad
\vec{P}_{\mathcal{R}} ^1= \frac{1}{3}a(T^{1})^4 \vec{\rm{D}}.
\end{equation}

For the two fluid electron-ion model, according to the Chapman-Enskog theory \cite{Chapman1990}, the zeroth order expansion $f_\alpha = g^\star$ gives the Euler RH equations
\begin{equation}\label{ap-zerothorder}
\begin{aligned}
	&\partial_{t}\rho+\partial_{\vec{x}}\cdot(\rho\vec{U})=0,\\
	&\partial_{t}(\rho\vec{U})+\partial_{\vec{x}}\cdot(\rho\vec{U}\vec{U}+p \vec{\rm{D}})= -\frac{1}{c} \int \vec{\Omega} S_{\mathcal{R}} {\rm d} \vec{\Omega},\\
	&\partial_{t}E+\partial_{\vec{x}}\cdot\left((E+p)\vec{U}\right)= -\frac{1}{\epsilon} \int S_{\mathcal{R}} {\rm d} \vec{\Omega},\\
\end{aligned}
\end{equation}
where total density $\rho$, total number density $n$, total momentum $\vec{U}$, total energy $E$ and total pressure $p$ satisfy
\begin{equation}\label{multispecies property1}
\begin{aligned}
	\rho=\sum_{\alpha=\mathcal{E}}^{N}\rho_{\alpha}, \quad n=\sum_{\alpha=\mathcal{E}}^{N}n_{\alpha},\quad
	\rho{\vec{U}}=\sum_{\alpha=\mathcal{E}}^{N}\rho_{\alpha}{\vec{U}}_{\alpha},\quad E=\sum_{\alpha=\mathcal{E}}^{N}E_{\alpha},\quad
	{p} = \frac{2}{3n k_{B}}(E - \frac{1}{2}\rho\Vec{U}^2),
\end{aligned}
\end{equation}
where the energy of the species is given by
\begin{equation}\label{eq:E-alpha}
E_\alpha = \frac{1}{2}\rho_\alpha \vec{U}_{\alpha}^2 + \frac{3\rho_\alpha k_{B}T_{\alpha}}{2 m_\alpha}.
\end{equation}

Similarly, expanding the variables as a power series of $\epsilon$,
\begin{equation}\label{aap-expand}
\rho = \sum^{\infty}_{i = 0} \rho^{i}\epsilon^{i},
\quad
\vec{U} = \sum^{\infty}_{i = 0} \vec{U}^{i}\epsilon^{i},
\quad
T = \sum^{\infty}_{i = 0} T^{i}\epsilon^{i},
\end{equation}
and combining with the above analysis of radiation transfer model, the $\mathcal{O}(\epsilon^0)$ terms in the Eq.\eqref{ap-zerothorder} is
\begin{equation}\label{ap-full-0}
\begin{aligned}
	&\partial_{t}\rho^0 + \partial_{\vec{x}}\cdot(\rho^0\vec{U}^0)=0,\\
	&\partial_{t}(\rho^0\vec{U}^0)+\partial_{\vec{x}}\cdot(\rho^0\vec{U}^0\vec{U}^0+p^0\vec{\rm{D}} + p_{\mathcal{R}}^0\vec{\rm{D}})= \vec{0},\\
	&\partial_{t}(E^0 + E_{\mathcal{R}}^0)+\partial_{\vec{x}}\cdot \big(\vec{U}^0 ( E^0+p^0+\frac{4}{3} E_{\mathcal{R}}^0 ) + \vec{F}_{\mathcal{R}}^1\big) =  0.
\end{aligned}
\end{equation}
By substituting Eq.\eqref{rad-ap1st} into Eq.\eqref{ap-full-0}, in continuum limit under the condition $\tau \ll 1, \epsilon \to 0$, the kinetic model can reduce to the following radiation plasma hydrodynamic equations,
\begin{equation}\label{ap-macro}
\begin{aligned}
	&\partial_{t}\rho + \partial_{\vec{x}}\cdot(\rho\vec{U})=0,\\
	&\partial_{t}(\rho\vec{U})+\partial_{\vec{x}}\cdot(\rho\vec{U}\vec{U}+p\vec{\rm{D}} + p_{\mathcal{R}}\vec{\rm{D}})= \vec{0},\\
	&\partial_{t}(E + E_{\mathcal{R}})+\partial_{\vec{x}}\cdot \big(\vec{U} ( E+p+\frac{4}{3}E_{\mathcal{R}} ) \big) =  \partial_{\vec{x}} \cdot (D_{\mathcal{R}} \partial_{\vec{x}} E_{\mathcal{R}}),
\end{aligned}
\end{equation}
where total density $\rho$, total momentum $\vec{U}$, total energy $E$ and total pressure $p$ are defined in Eq.\eqref{multispecies property1}. And $E_{\mathcal{R}} = acT^{4}_{\mathcal{R}}$ is radiative energy, the diffusion coefficient $D_{\mathcal{R}}$ in the diffusion limit is defined by the single group Rosseland mean opacity $\sigma_{t}$ as $D_{\mathcal{R}} = {c}/{3\sigma_{t}}$.

\section{Gas-kinetic scheme and Unified gas-kinetic scheme}\label{sec:scheme}

This paper employs the gas-kinetic scheme (GKS) for ion and electron and the unified gas-kinetic scheme(UGKS) for radiation to solve the radiation plasma system with the proposed kinetic models. The GKS for multispecies mixture and the UGKS for radiative transfer equation will be introduced first. Subsequently, the numerical treatments and algorithms for solving the radiation plasma model will be discussed, respectively.

\subsection{Gas-kinetic scheme for ion and electron}
This section will present the construction of GKS with the kinetic model of a multispecies mixture. Within the finite volume framework, the update of the macroscopic variables $\vec{W}_\alpha$ of the species $\alpha$ in the finite control volume $i$ at a discrete time scale $\Delta t = t^{n+1}-t^{n}$ satisfies the fundamental conservation laws of particle evolution
\begin{equation}\label{eq:fvm-f}
	\vec{W}_{i,\alpha}^{n + 1}=\vec{W}_{i,\alpha}^n - \frac{\Delta t}{V_i} \sum\limits_{j \in N(i)} {\vec{F}}_{ij,\alpha} {\mathcal A}_{ij} +\vec{S}_{i,\alpha}.
\end{equation}
where $V_{i}$ is the volume of cell $i$, $\vec{W}_{i,\alpha}$ is the conservative variables of the species $\alpha$ of the cell $i$, i.e., the densities of mass $\rho_\alpha$, momentum $\rho_\alpha {\vec U}_{\alpha}$, and total energy $E_\alpha$.  $N(i)$ contains all the cell number of neighbor cells that is interface-adjacent with cell $i$ and cell $j$ is one of the neighbors. The interface between cells $i$ and $j$ is represented by the subscript $ij$. Hence, ${\mathcal A}_{ij}$ is referred to as the area of the interface $ij$. The source term $\vec{S}_{i,\alpha}$ accounts for the energy exchange caused by collisions between different species
\begin{equation} \label{eq:source-aap}
	\vec{S}_{i, \alpha} = \int_{0}^{\Delta t} \int \frac{g_{i,\alpha}^M-f_{i,\alpha}}{\tau_{i,\alpha}} \vec{\psi}
	{{\rm d}} \vec{u} {{\rm d}} t
	= \Delta t \vec{s}_{i,\alpha},
\end{equation}
where $\vec{s}$ can be expressed as
\begin{equation*}
	\vec{s}_\alpha = \left(
	0,
	\frac{
		(\rho_\alpha \vec{U}_{\alpha}^M)
		- (\rho_\alpha \vec{U}_\alpha)^\dagger}{\tau^\dagger_\alpha},
	\frac{(E_\alpha^M) -(E_\alpha)^\dagger}{\tau_\alpha^\dagger}
	\right)^T,
\end{equation*}
where the superscript $\dagger$ denotes the updated intermediate state with inclusion of the fluxes only
\begin{equation}\label{eq:update-mid}
	\vec{W}_{i,\alpha}^{\dagger} = {\vec W}_{i,\alpha}^{n}-\frac{\Delta t}{V_{i}} \sum\limits_{j \in N(i)}\vec{F}_{ij,\alpha} {\mathcal{A}}_{ij},
\end{equation}
and the $(\vec{U}^M)$ and $(E^M)$ can be calculated by Eq.~\eqref{aap-exchange} and Eq.~\eqref{eq:E-alpha}. The time-averaged  microscopic flux ${\mathcal F}_{ij,\alpha}$  and macroscopic flux $\vec{F}_{ij,\alpha}$ can be expressed as
\begin{equation}\label{eq:micro-flux}
	\mathcal{F}_{ij,\alpha} = \frac{1}{\Delta t} \int_0^{\Delta t} \vec{u} \cdot \vec{n}_{ij} f_{ij,\alpha}(t) {\rm d}t,
\end{equation}
and
\begin{equation}\label{eq:macro-flux-aap}
	\vec{F}_{ij,\alpha} = \int \mathcal{F}_{ij,\alpha} \vec{\psi} {\rm d} \vec{u}
\end{equation}
where $\vec{u}$ is the microscopic particle velocity, and $\vec{n}_{ij}$ is the normal vector of the interface $ij$, $\vec{\psi} = (1, \vec{u}, \frac12 \vec{u}^2)^T$, $f_{ij,\alpha}(t)$ is the time-dependent gas distribution function on the interface constructed by the integral solution of the kinetic model Eq.~\eqref{eq:kinetic-model}
\begin{equation*}
	f_\alpha(\vec{x},t)={\frac{1}{\tau_\alpha}}\int_{0}^{t}g_\alpha(\vec{x}^{\prime},t^{\prime})e^{-(t-t^{\prime})/\tau_\alpha}{\rm d} t^{\prime}+e^{-t/\tau}f^0_{\alpha}(\vec{x}-\vec{u}t),
\end{equation*}
where $f^0_\alpha({\vec x})$ is the initial distribution function of the species $\alpha$ at the beginning of each step $t_n$, and $g_\alpha(\vec{x}, t)$ is the equilibrium state distributed in space and time around $\vec{x}$ and $t$. The integral solution couples the particle-free transport and collisions in the gas evolution process. To achieve the second-order accuracy, the Taylor expansion is applied to the equilibrium state $g_\alpha$ and first-order Chapman–Enskog expansion is applied to initial distribution function $f_\alpha^0$  to recover macroscopic transport coefficients
\begin{equation}\label{eq:Taylor}
	\begin{aligned}
		&g_\alpha(\vec{x},t) = g_\alpha^0
		+ \vec{x} \cdot \frac{\partial g_{\alpha}}{\partial \vec{x}}
		+ \frac{\partial g_{\alpha}}{\partial t} t, \\
		&f_{\alpha}^{0}(\vec{x}) =
		g_{\alpha}^{l,r} + \vec{x}\cdot \frac{\partial g_{\alpha}^{l,r}}{\partial \vec{x}} -\tau\left(\vec{u} \cdot \frac{\partial g_{\alpha}^{l,r}}{\partial \vec{x}}+\frac{\partial g_{\alpha}^{l,r}}{\partial t}\right),
	\end{aligned}
\end{equation}
where $g_{\alpha}^{l,r}$ is the distribution function constructed by distribution functions $g_\alpha^l$ and $g_\alpha^r$ interpolated from cell centers to the left and right sides of the interface
\begin{equation*}
	g_{\alpha}^{l,r} = g_{\alpha}^l H\left[\bar{u}_{ij}\right]
	+ g_\alpha^r\left(1-H\left[\bar{u}_{ij}\right]\right),
\end{equation*}
where $\bar{u}_{ij}=\vec{u}\cdot\vec{n}_{ij}$ is the particle velocity projected on the normal direction of the cell interface $\vec{n}_{ij}$, and $H[x]$ is the Heaviside function. The equilibrium state $g^0_{\alpha}$ at the cell interface is computed from the colliding particles from both sides of the cell interface
\begin{equation}
	\vec{W}^0_\alpha
	= \int g^0_\alpha \vec{\psi} {\rm d} \vec{u}
	= \int g^{l,r}_\alpha \vec{\psi} {\rm d} \vec{u},
\end{equation}
and the spatial and temporal derivatives of the equilibrium state can be obtained by
\begin{equation}
	\begin{aligned}
		& \int \frac{\partial g_\alpha}{\partial \vec{x}} \vec{\psi}  {\rm d}  \vec{u}
		= \frac{\partial \vec{W}_\alpha}{\partial \vec{x}}, \\
		& \int \frac{\partial g_\alpha}{\partial t} \vec{\psi}  {\rm d}  \vec{u}
		= - \int \vec{u} \cdot \frac{\partial g_\alpha}{\partial \vec{x}} \vec{\psi}  {\rm d}  \vec{u},
	\end{aligned}
\end{equation}
where the spatial derivatives of the conservative variables ${\partial \vec{W}_\alpha}/{\partial \vec{x}}$ can be obtained from reconstruction. The temporal gradients $\partial_t g_{\alpha}^{l,r}$ are also evaluated from $\partial \vec{W}_{\alpha}^{l}/\partial{\vec t}$ and $\partial \vec{W}_{\alpha}^{r}/\partial{\vec t}$ by employing the compatibility condition
\begin{equation*}
	\begin{aligned}
		\frac{\partial\vec{W}^{l}_{\alpha}}{\partial t}
		= - \int \vec{u} \cdot \frac{\partial g^l_{\alpha}}{\partial \vec{x}} \vec{\psi} {\rm d}\vec{u}, \\
		\frac{\partial\vec{W}^{r}_{\alpha}}{\partial t}
		= - \int \vec{u} \cdot \frac{\partial g^r_{\alpha}}{\partial \vec{x}} \vec{\psi} {\rm d}\vec{u}.
	\end{aligned}
\end{equation*}

Substituting Eq~\eqref{eq:Taylor} into Eq~\eqref{eq:micro-flux}, we can get the time-averaged microscopic flux
\begin{equation}\label{eq:micro-flux-aap}
	\mathcal{F}_{ij,\alpha}=\vec{u} \cdot \vec{n}_{ij}
	\left(
	C_{1}g^0_{\alpha}+C_{2} \frac{\partial g_\alpha}{\partial \vec{x}} \cdot \vec{u}+C_{3} \frac{\partial g_\alpha}{\partial t}
	\right)
	+ \vec{u} \cdot \vec{n}_{ij}
	\left(
	C_4 g_{\alpha}^{l,r}
	+ C_5 \vec{u} \cdot \frac{\partial g_{\alpha}^{l,r}}{\partial \vec{x}}
	+ C_6 \frac{\partial g_{\alpha}^{l,r}}{\partial t}
	\right)
\end{equation}
with time coefficients
\begin{equation*}
	\begin{aligned}
		C_1 &= 1 - \frac{\tau}{\Delta t} \left( 1 - e^{-\Delta t / \tau} \right) , \\
		C_2 &= -\tau + \frac{2\tau^2}{\Delta t} - e^{-\Delta t / \tau} \left( \frac{2\tau^2}{\Delta t} + \tau\right) ,\\
		C_3 &=  \frac12 \Delta t - \tau + \frac{\tau^2}{\Delta t} \left( 1 - e^{-\Delta t / \tau} \right) , \\
		C_4 &= \frac{\tau}{\Delta t} \left(1 - e^{-\Delta t / \tau}\right), \\
		C_5 &= \tau  e^{-\Delta t / \tau} - \frac{2\tau^2}{\Delta t}(1 -  e^{-\Delta t / \tau}),\\
		C_6 &= - \frac{2\tau^2}{\Delta t}(1 -  e^{-\Delta t / \tau}).
	\end{aligned}
\end{equation*}

\subsection{Unified gas-kinetic scheme for radiation}
Under the framework of the finite volume method, the discretized equation of radiation intensity in cell $i$ can be written as
\begin{equation}\label{eq:I}
	I_i^{n+1}=I_i^n-\frac{\Delta t}{V_i} \sum_{j \in N(i)} \mathcal{F}_{i j, \mathcal{R}} \mathcal{A}_{i j}+\int_0^{\Delta t}S_{\mathcal{R}} {\rm d} t,
\end{equation}
where the $\mathcal{F}_{ij,\mathcal{R}}$ is the numerical flux across the interface
\begin{equation}
	\mathcal{F}_{ij,\mathcal{R}}=\frac{c}{\epsilon\Delta t}\int_{0}^{\Delta t} \vec{\Omega}\cdot \bigg(\vec{n}_{ij} I_{ij}(t) + \epsilon\theta\vec{\beta} I_{ij}(t) \bigg){\rm d}t = \mathcal{F}_{ij,\mathcal{R}}^{1} + \mathcal{F}_{ij,\mathcal{R}}^{2},\\
\end{equation}
constructed by the solution $I_{ij}(t)$ that is obtained by the integral solution of the radiative transfer equation
\begin{equation}
	\begin{aligned}
		I(\vec{x}_0,t)= & e^{-\nu(t)}I^{0}\Big(\vec{x}_0-\frac{c}{\epsilon}\vec{\Omega}t\Big) +\int_{0}^{\Delta t}
		e^{-\nu(t-t^{\prime})}\frac{c}{2\pi\epsilon}\big((\frac{\sigma_{t}}{\epsilon} -  	\epsilon\sigma_{s})\phi + \sigma_{t}\epsilon c E_{\mathcal{R}}\big)(\vec{x}_0-\frac{c}{\epsilon}\vec{\Omega}(t - t^{\prime}), t^{\prime}){\rm d} t^{\prime} \\
		&+ \int_{0}^{\Delta t} e^{-\nu(t-t^{\prime})} c \frac{\partial}{\partial \vec{x}} \cdot (\theta \vec{\beta} I ){\rm d} t^{\prime}
	\end{aligned}
\end{equation}
where $I^0$ denotes the radiation intensity at the initial state of the current step, and $\phi$ is the local equilibrium state of radiation intensity. One can get  $\phi$ and $I^0$ in second-order accuracy by Taylor expansion
\begin{equation}
	\begin{aligned}
		\phi(\vec{x},t)&=\phi^{0}
		+ \frac{\partial \phi}{\partial \vec{x}}\cdot \vec{x}
		+ \frac{\partial \phi}{\partial t} t, \\
		I^{0}(\vec{x})&=I^{0}
		+ \frac{\partial I}{\partial \vec{x}}\cdot \vec{x}.
	\end{aligned}
\end{equation}
By integrating the distribution function, we can get the microscopic flux over a time step is
\begin{equation}\label{eq:micro-flux-rad}
	\begin{aligned}
		\mathcal{F}_{ij,\mathcal{R}}^{1}
		=
		& \vec{\Omega}\cdot \vec{n}_{ij}\left(
		A_1 I^0 + A_2 \frac{\partial I}{\partial \vec{x}} \cdot \vec{\Omega}
		\right)
		+ \vec{\Omega}\cdot \vec{n}_{ij} \left(
		A_3 \phi^{0}
		+ A_4 \frac{\partial \phi}{\partial \vec{x}}\cdot \vec{\Omega}
		+ A_5 \frac{\partial \phi}{\partial t}
		\right)\\
		&+\vec{\Omega}\cdot \vec{n}_{ij} \left(  A_6 ( \epsilon \frac{\partial}{\partial \vec{x}} \cdot (\theta \vec{\beta} I))\right)
	\end{aligned}
\end{equation}
with the time coefficients
\begin{equation}
	\begin{aligned}
		&A_1=\frac{c}{\epsilon\Delta t\nu}\left[1-e^{-\nu\Delta t}\right],\\
		&A_2=-\frac{c^{2}}{\epsilon^{2}\nu^{2}\Delta t}\left[1-e^{-\nu\Delta t}-\nu\Delta t e^{-\nu\Delta t}\right],\\
		&A_3=\frac{c^{2}(\frac{\sigma_t}{\epsilon} - \epsilon\sigma_s)}{2\pi\Delta t\epsilon^{2}\nu}
		\left[\Delta t-\frac{1}{\nu}\left(1-e^{-\nu\Delta t}\right)\right],\\
		&A_4=-\frac{c^{3}(\frac{\sigma_t}{\epsilon} - \epsilon\sigma_s)}{2\pi\,\Delta t\epsilon^{3}\nu^{2}}
		\left[\Delta t\left(1+e^{-\nu\Delta t}\right)-\displaystyle\frac{2}{\nu}\left(1-e^{-\nu\Delta t}\right)\right],\\
		&A_5=\frac{c^{2}(\frac{\sigma_t}{\epsilon} - \epsilon\sigma_s)}{2\pi\epsilon^{2}\nu^{3}\Delta t}
		\left[
		1-e^{-\nu\Delta t}-\nu\Delta t e^{-\nu\Delta t}-\frac{1}{2}(\nu\Delta t)^{2}
		\right],\\
		&A_6 = \frac{c^2}{\epsilon^2\Delta t\nu}\left[\Delta t-\frac{1}{\nu}\left(1-e^{-\nu\Delta t}\right)\right],
	\end{aligned}
\end{equation}
where $\nu = {c\sigma}/{\epsilon^2}$. And for $\mathcal{F}_{ij,\mathcal{R}}^{2}$, the upwind side cell center value is used by the sign of the fluid velocity $\vec{U}$ at the boundary,
\begin{equation}
	\mathcal{F}_{ij,\mathcal{R}}^{2} = \left\{
	\begin{aligned}
		\theta \vec{U}_{j} \cdot \vec{n}_{ij}  I_{j}&, &\vec{n}_{ij} \cdot(\vec{U}_{i} + \vec{U}_{j}) < 0 \\
		\theta \vec{U}_{i} \cdot \vec{n}_{ij}  I_{i}&, &\vec{n}_{ij} \cdot(\vec{U}_{i} + \vec{U}_{j}) > 0
	\end{aligned}	\right.
\end{equation}

And thus the construction of the numerical scheme is completed.
For the radiation plasma kinetic model used in this paper, the microscopic radiation intensity will be macroscopically reflected by radiation energy $E_{\mathcal{R}}$ by taking moments. Therefore, the discretized macroscopic equation can be derived by taking moments of Eq.~\eqref{eq:I}
\begin{equation}
	(E_{\mathcal{R}})_i^{n+1}=(E_{\mathcal{R}})_i^n-\frac{\Delta t}{V_i} \sum_{j \in N(i)} {F}_{ij,\mathcal{R}} \mathcal{A}_{i j} + \int_{0}^{\Delta t} \int S_{\mathcal{R}} {\rm d} \vec{\Omega} {\rm d} t,
\end{equation}
where $F_{ij, \mathcal{R}}$ denotes the macroscopic flux for radiation energy of the cell interface $ij$
\begin{equation}\label{eq:macro-flux-rad}
	F_{ij,\mathcal{R}} = \int \mathcal{F}_{ij,\mathcal{R}} {\rm d} \vec{\Omega}.
\end{equation}

\subsection{GKS-UGKS for radiation plasma system}
We build a numerical simulation method for a radiation plasma system with the electron-ion-photon model. The discretized microscopic equations of the election-ion-photon model Eq.~\eqref{eq:kinetic-model} is
\begin{equation}\label{eq:update-micro-eip}
	\left[\begin{array}{l}
		f_\alpha \\
		I
	\end{array}\right]_i^{n+1}=\left[\begin{array}{l}
		f_\alpha \\
		I
	\end{array}\right]_i^n-\frac{\Delta t}{V_i} \sum_{j \in N(i)}\left[\begin{array}{c}
		\mathcal{F}_{i j, \alpha} \\
		\mathcal{F}_{i j, \mathcal{R}}
	\end{array}\right] \mathcal{A}_{i j}+\int_0^{\Delta t}\left[\begin{array}{l}
		\left(g_{i, \alpha}^M-f_{i, \alpha}\right) / \tau_{i, \alpha} \\
		S_{\mathcal{R}}
	\end{array}\right] \mathrm{d} t,
\end{equation}
where the species $\alpha \in \{\mathcal{E}, \mathcal{I}\}$. The details about the construction of flux and the relaxation terms of the microscopic distribution functions and radiation intensity coordinate with the GKS for multispecies mixture and the UGKS for radiation, respectively. The discretized macroscopic equation can be written as
\begin{equation}\label{eq:update-macro-eip}
	\left[\begin{array}{c}
		\vec{W}_\alpha \\
		E_\mathcal{R}
	\end{array}\right]_i^{n+1}=\left[\begin{array}{c}
		\vec{W}_\alpha \\
		E_\mathcal{R}
	\end{array}\right]_i^n-\frac{\Delta t}{V_i} \sum_{j \in N(i)}\left[\begin{array}{c}
		\vec{F}_{i j, \alpha} \\
		F_{i j, \mathcal{R}}
	\end{array}\right] \mathcal{A}_{i j}+\left[\begin{array}{c}
		\vec{S}_{i, \alpha} \\
		0
	\end{array}\right]+\left[\begin{array}{c}
		\vec{S}_{i, \alpha \mathcal{R}} \\
		S_{i, \mathcal{R} \alpha}
	\end{array}\right],
\end{equation}
where the source term $\vec{S}_{i,\alpha \mathcal{R}}$ and $S_{i, \mathcal{R} \alpha}$ denotes the relaxation process between radiation and ion or electron with
\begin{equation}\label{eq:source-eip}
	\begin{aligned}
		& \vec{S}_{{\mathcal {I}\mathcal{R} }}=(0, \vec{0}, 0)^T,\\
		& \vec{S}_{\mathcal{ER}}=\left(0,\Delta t(\rho_\mathcal{E}\vec{U}_\mathcal{E}^M - \rho_\mathcal{E}\vec{U}_\mathcal{E})/\tau_{\mathcal{E}}, \Delta t (E_{\mathcal{E}}^{M}-E_{\mathcal{E}})/\tau_{\mathcal{E}}\right)^T, \\
		& {S}_{\mathcal{R} \alpha}=S_{\mathcal{R} \mathcal{E}}=\Delta t \bigg( \int S_{\mathcal{R}} {\rm d} \vec{\Omega}\bigg).
	\end{aligned}
\end{equation}
The algorithms for one time step evolution ($t^n \to t^{n+1}$) of the GKS-UGKS can be summarized as follows:\\
{\bf1.} Obtain the macroscopic numerical fluxes of multispecies mixture Eq.~\eqref{eq:macro-flux-aap} and both the microscopic and macroscopic numerical fluxes of radiation transfer Eq.~\eqref{eq:micro-flux-rad} and Eq.~\eqref{eq:macro-flux-rad}. \\
{\bf2.} Obtain the first updated intermediate macroscopic variables with the fluxes only by Eq.~\eqref{eq:update-mid}. \\
{\bf3.} Obtain the source term between the electron and ion by multispecies mixture using Eq.~\eqref{eq:source-aap}. \\
{\bf4.} Obtain the second updated intermediate state by updating the macroscopic variables with the inclusion of source term caused by multispecies mixture using Eq.~\eqref{eq:update-mid}. \\
{\bf5.} Obtain the momentum and energy exchange terms between electron and radiation caused by radiative background field Eq.~\eqref{eq:source-eip}. \\
{\bf6.} Update the macroscopic variables by Eq.~\eqref{eq:update-macro-eip} and microscopic distribution function by Eq.~\eqref{eq:update-micro-eip} in the $n+1$ time.

\section{Numerical cases}\label{sec:case}

\subsection{Radiative transfer in kinetic and diffusion regime}
\begin{figure}[H]
	\centering
	\subfloat[]{\includegraphics[width=0.4\textwidth]{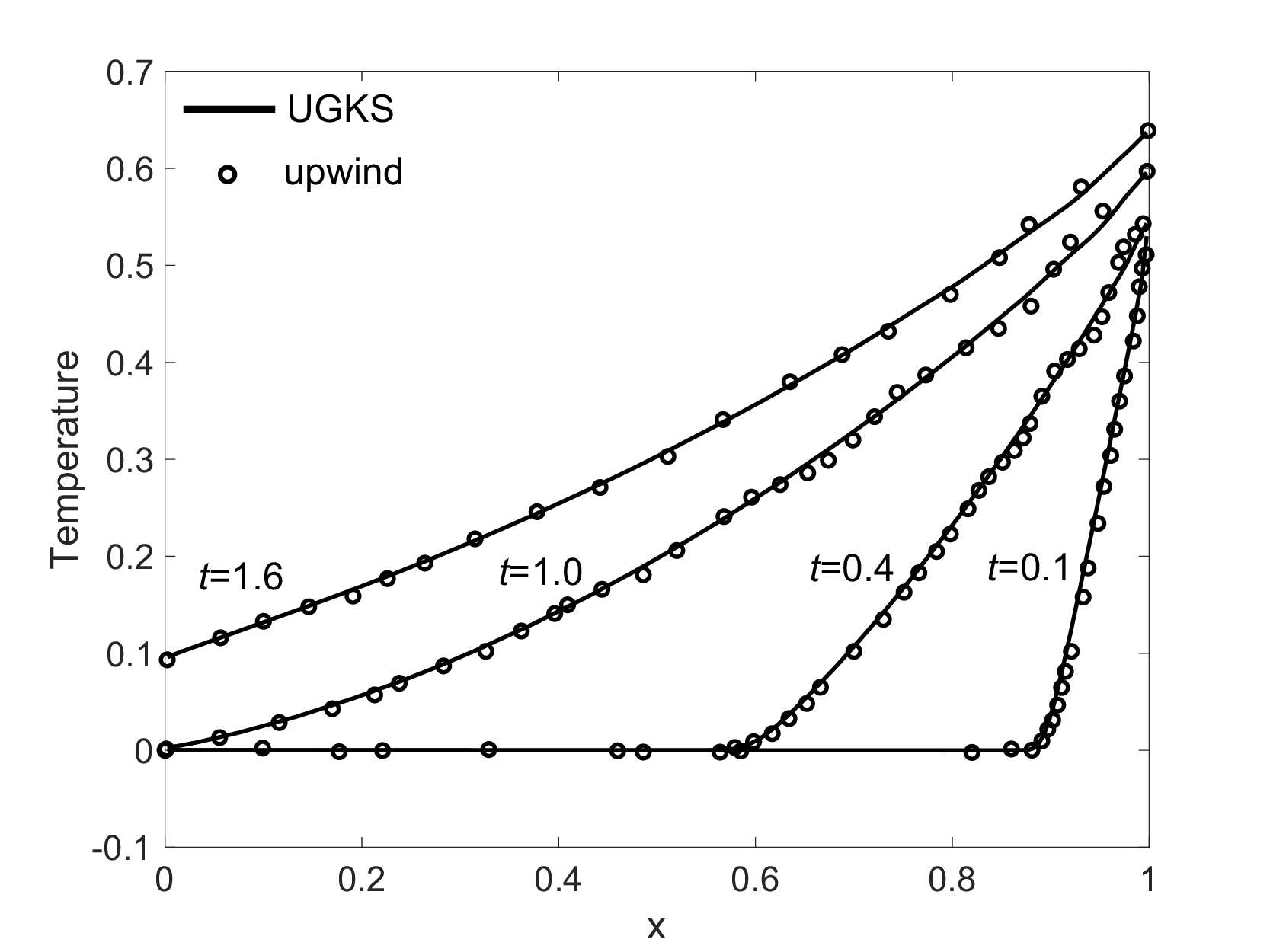}}
	\subfloat[]{\includegraphics[width=0.4\textwidth]{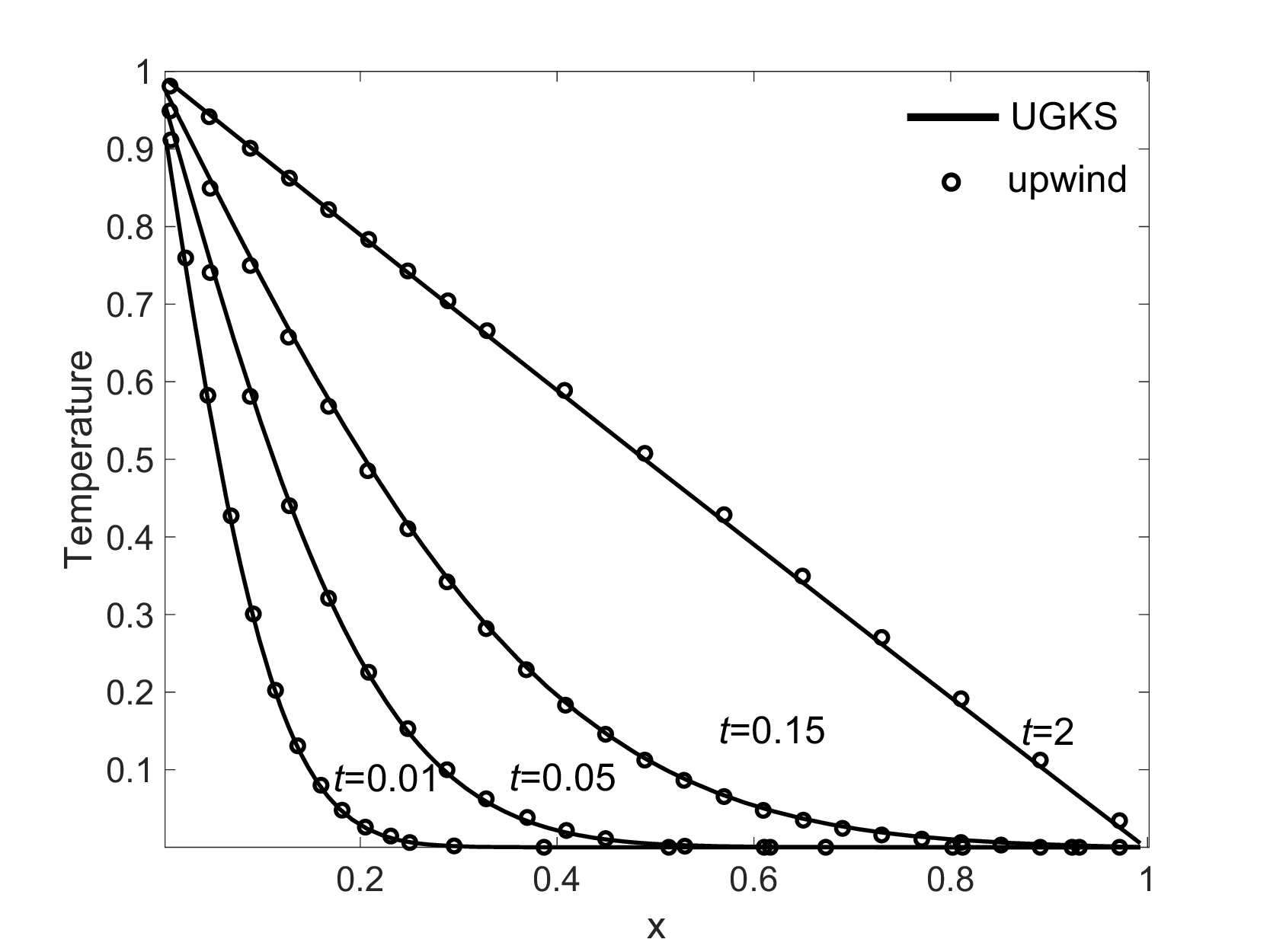}}
	\caption{Distributions of the radiative transfer temperature in (a) the kinetic regime at times at time $t = 0.1, 0.4, 1.0, 1.6$, and (b) in the diffusion regime at time $t = 0.01, 0.05, 0.15, 2.0$, solved by UGKS compared with the upwind explicit discretization of the radiation rexlation model \cite{mieussens2013}.}
	\label{fig:rad-1D}
\end{figure}

The radiative transfer under optically thick and optically thin regimes will be computed to validate the capability of UGKS for multiscale simulation in radiation transport. The exchange coefficient between radiation and electron is set as zero. The non-dimensional initial condition for radiative transfer in the kinetic regime is
\begin{equation}
	I(x)=0, x \in(0,1), \quad I(0)=0, I(1)=1, \quad \epsilon=1,
\end{equation}
and in the diffusion regime
\begin{equation}
	I(x)=0, x \in(0,1), \quad I(0)=1, I(1)=0, \quad \epsilon=10^{-8}.
\end{equation}
The physical space is discretized with 200 uniform meshes and the velocity space is with 30 meshes by midpoint rule. The range of velocity space is $\Omega_x\in(-1,1)$. Fig.~\ref{fig:rad-1D}(a) shows the radiation temperatures in the kinetic regime at $t = 0.1, 0.4, 1.0, 1.6$. Fig.~\ref{fig:rad-1D}(b) is in the diffusion regime at $t = 0.01, 0.05, 0.15, 2.0$. The results are compared with the reference solution solved by upwind explicit discretization of the radiation relaxation model from \cite{mieussens2013}. Good agreements in both optically thin and thick limits exhibit accuracy when no interaction between electron and radiation happens.

\subsection{Marshak wave}
To further evaluate the capability of the current scheme in capturing the radiative equilibrium diffusion solution under conditions of optically thick limit, we impose a zero fluid velocity and maintain energy exchange between the material and radiation. The absorption/emission coefficient is $\sigma = 100/T^{3}$. The initial condition of material temperature is $ T_{\mathcal{E}} = 10^{-4}$. And the initial condition for radiation field is $ I(x)=0, x \in(0,1), I(0)=1, I(1)=0,$. The physical space is discretized with 200 uniform meshes and the velocity space is with 30 meshes by midpoint rule. The range of velocity space is $\Omega_x\in(-1,1)$.
\begin{figure}[H]
	\centering
	\includegraphics[width=0.5\textwidth]{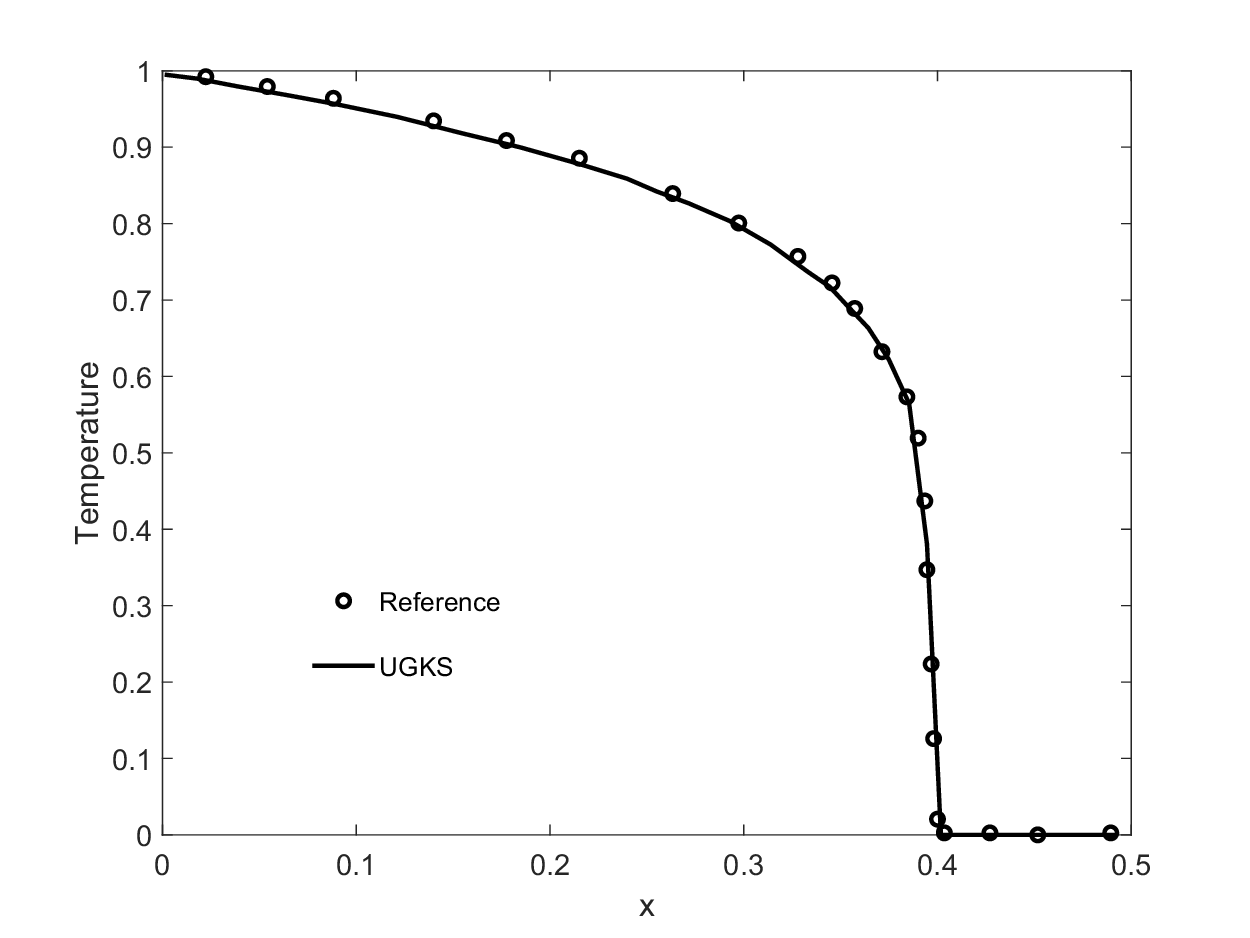}
	\caption{Marshak wave case whit $\sigma = 100/T^{3}$, solved by UGKS with the kinetic model compared with the result solved by diffusion limit solution \cite{sun2015gray}.}
	\label{fig:rad-gray}
\end{figure}

\subsection{Radiative shock for $\rm{Ma} = 1.5$ and $\rm{Ma} = 3.0$}
\begin{figure}[H]
	\centering
	\subfloat[]{\includegraphics[width=0.33\textwidth]{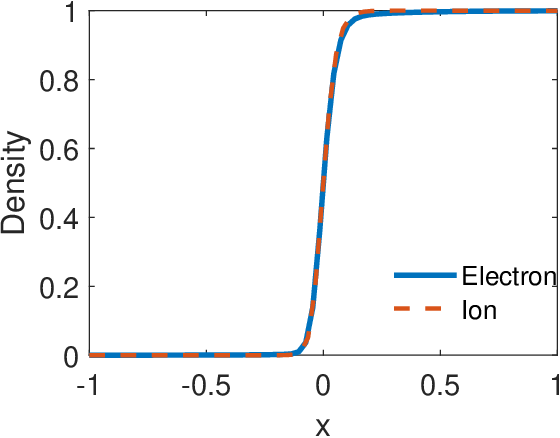}}
	\subfloat[]{\includegraphics[width=0.33\textwidth]{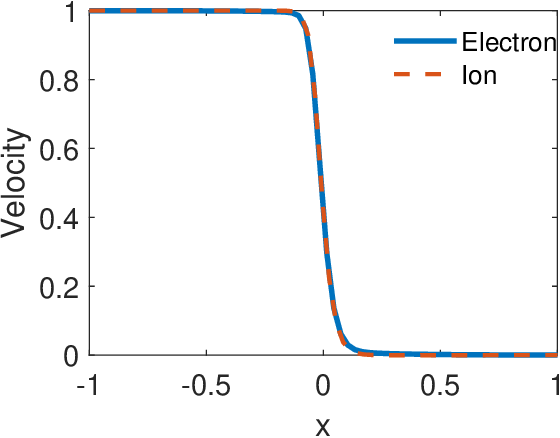}}
	\subfloat[]{\includegraphics[width=0.33\textwidth]{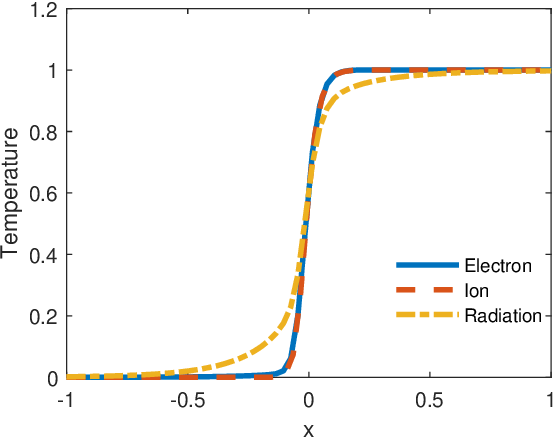}}
	\caption{Distributions of the radiative shock (a) density, (b) velocity, and (c) temperature with Mach number $\rm{Ma} = 1.5$ and mass ratio $m_\mathcal{E}/m_\mathcal{I} = 1.0$.}
	\label{fig:radshock-1.5}
\end{figure}

\begin{figure}[H]
	\centering
	\subfloat[]{\includegraphics[width=0.33\textwidth]{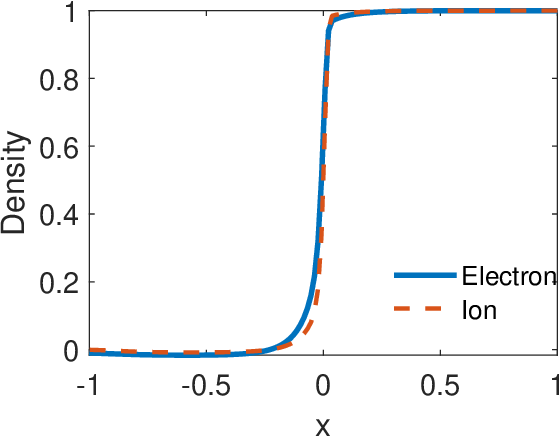}}
	\subfloat[]{\includegraphics[width=0.33\textwidth]{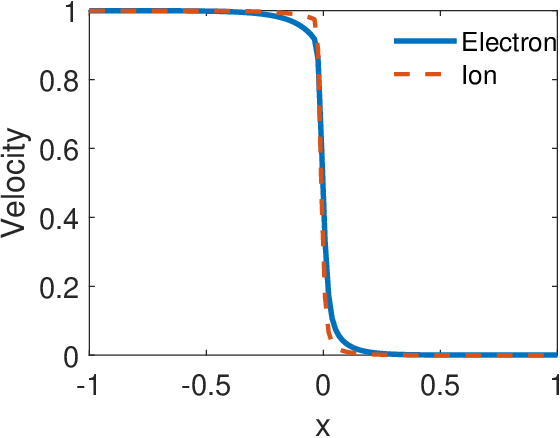}}
	\subfloat[]{\includegraphics[width=0.33\textwidth]{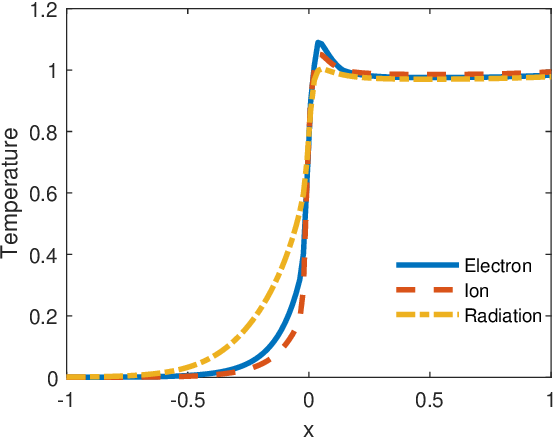}}
	\caption{Distributions of the radiative shock (a) density, (b) velocity, and (c) temperature with Mach number $\rm{Ma} = 3.0$ and mass ratio $m_\mathcal{E}/m_\mathcal{I} = 1.0$.}
	\label{fig:radshock-3.0}
\end{figure}

The newly developed multiscale method will be tested by two radiative shock cases, which are presented in \cite{bolding2017second} and \cite{lowrie2008radiative}. The radiation intensity is initialized according to the electron temperature. The physical space is discretized with 200 uniform meshes and the velocity space is with 30 meshes by midpoint rule. The range of velocity space is $\Omega_x\in(-1,1)$. Fig.~\ref{fig:radshock-1.5} shows the density velocity and temperature of electron and ion and also radiation temperature for the Mach number $\rm{Ma} = 1.5$ and mass ratio $m_\mathcal{E}/m_\mathcal{I} = 1.0$.  In the numerical solution, we observe a discontinuity in the fluid temperature due to the hydrodynamic shock, and the maximum temperature is bounded by the far-downstream
temperature. This matches with the results in \cite{lowrie2008radiative,bolding2017second}. Fig.~\ref{fig:radshock-3.0} shows the density velocity and temperature of electron, ion, and also radiation temperature for the $\rm{Ma} = 3.0$ and mass ratio $m_\mathcal{E}/m_\mathcal{I} = 1.0$.  For the strong shock, the material temperature reaches its maximum at the post-shock state, this point is called the Zel’dovich spike. In this case, both hydrodynamic shock and Zel’dovich spike appear which matches with the result \cite{lowrie2008radiative,bolding2017second}.

\subsection{3T-Double Lax shock tube problem}
The non-dimensional initial condition is
\begin{equation*}
	\left\{\begin{array}{l l l l}
		{\rho=0.415,\quad u=0.698,\quad p=1.176, \quad -1\leq x\leq-0.5}\\
		{\rho=0.5,  \qquad u=0,   ~~ \qquad p =0.19, \quad -0.5\leq x\leq0.5}\\
		{\rho=0.415,\quad u=0.698,\quad p=1.176, \quad 0.5\leq x\leq1}
	\end{array}\right.
\end{equation*}
with
\begin{equation*}
	\rho = \rho_{\mathcal{E}} = \rho_{\mathcal{I}}, \quad
	U = U_{\mathcal{E}}, \quad
	p = p_{\mathcal{E}} = p_{\mathcal{I}},
\end{equation*}
The election, ion, and radiation are assumed to be a hard-sphere collision model. The the molecular diameter $d = d_\mathcal{E} = d_\mathcal{I} $ and number density $n = n_\mathcal{E} + n_\mathcal{I} $ are set for the defination of Knudsen number. The physical space with the reference length for the definition of Knudsen numerical is $L_{ref} = 1$, discretized by 200 uniform meshes. The velocity space is discretized by 100 meshes using the midpoint rule with the range $u_\alpha \in (-7,7)$ according to the most probable speed of each species $\alpha$. The results at the time $t = 1$ are investigated.
The distributions of density, velocity, electron temperature, ion temperature, and radiation temperature are quantitively compared with the results of three temperature (3T) radiation hydrodynamics (RH) \cite{cheng2024}.

\begin{figure}[H]
	\centering
	\subfloat[]{\includegraphics[width=0.33\textwidth]{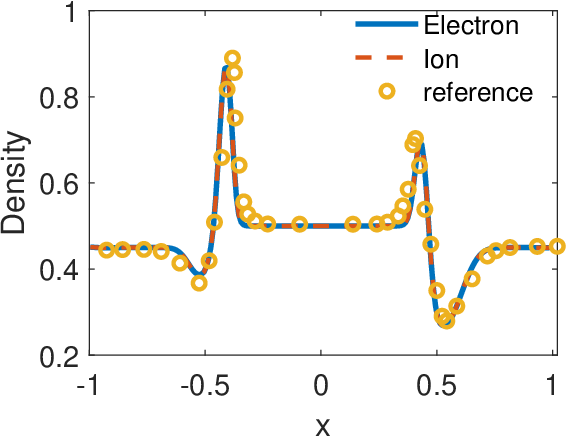}}
	\subfloat[]{\includegraphics[width=0.33\textwidth]{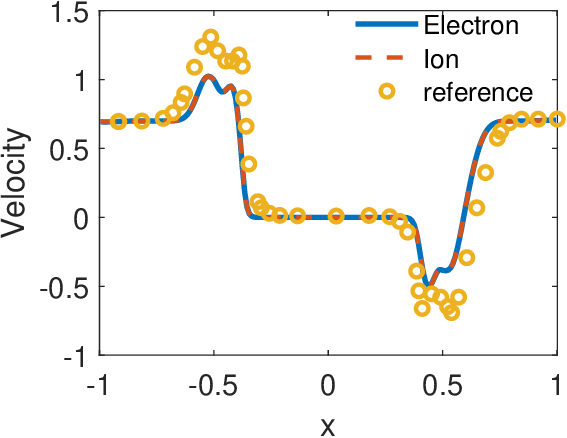}}
	\subfloat[]{\includegraphics[width=0.33\textwidth]{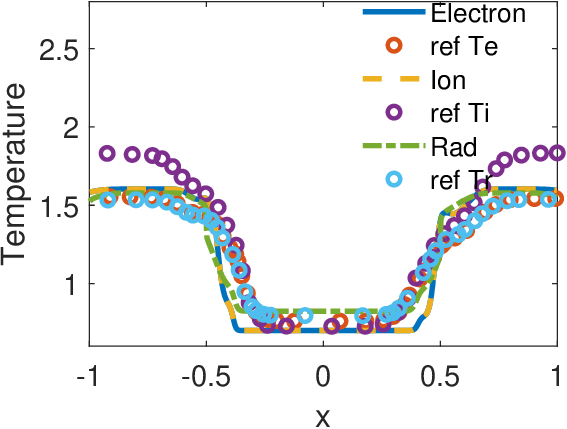}} \\
	\subfloat[]{\includegraphics[width=0.33\textwidth]{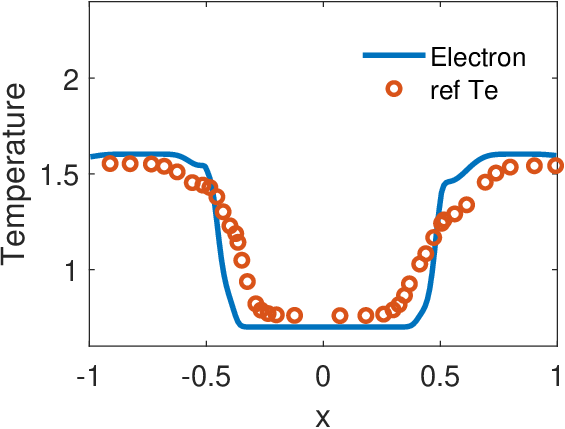}}
	\subfloat[]{\includegraphics[width=0.33\textwidth]{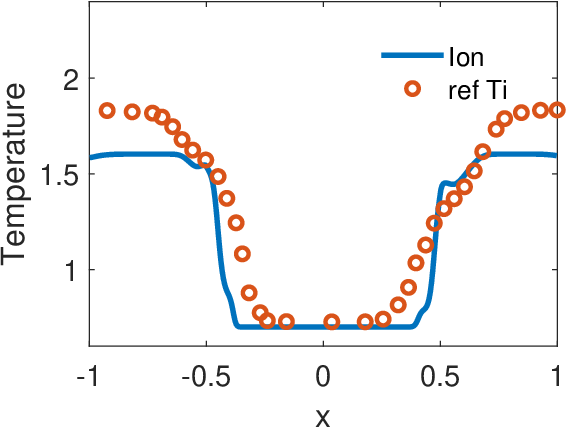}}
	\subfloat[]{\includegraphics[width=0.33\textwidth]{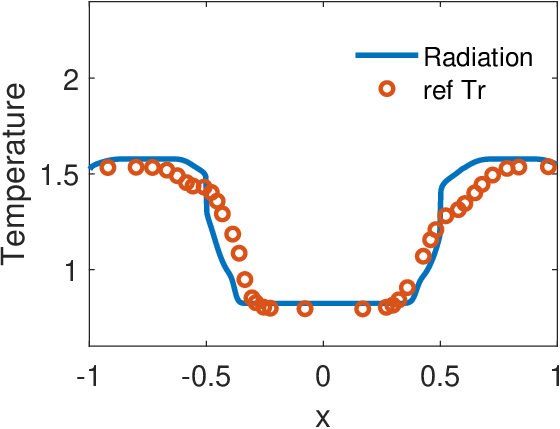}}
	\caption{Two interacting blast wave in the transition flow regime at ${\rm Kn} = 10^{-5}$ with the mass ratio $m_\mathcal{I}/m_\mathcal{E} = 1.0$ and number density ratio $n_\mathcal{I}/n_\mathcal{E} = 1.0$ of ion to election. Distributions of (a) densities, (b) velocities, (c) temperatures of election, ion, and radiation, (d) temperature of electron, (e) temperature of ion, and (e) temperature of radiation. Compared with three-temperature radiation hydrodynamics (RH-3T) solved by \cite{cheng2024}.}
	\label{fig:blast-kn1e-5}
\end{figure}

Fig.~\ref{fig:blast-kn1e-5} shows that our result of density matches good with the result from \cite{cheng2024}. And due to the momentum exchanges with the radiation field, the velocity is with same shape but a bit less compared with the result from \cite{cheng2024}. Temperature varies from the result in \cite{cheng2024}. In our case, the temperature exchange is obtained by the integral of the radiative intensity while \cite{cheng2024} use the hydrodynamic model and the exchange is obtained by a parameter $\omega$.

\subsection{Sedov-Taylor one point blast wave}
\begin{figure}[H]
	\centering
	\subfloat[]{\includegraphics[width=0.33\textwidth]{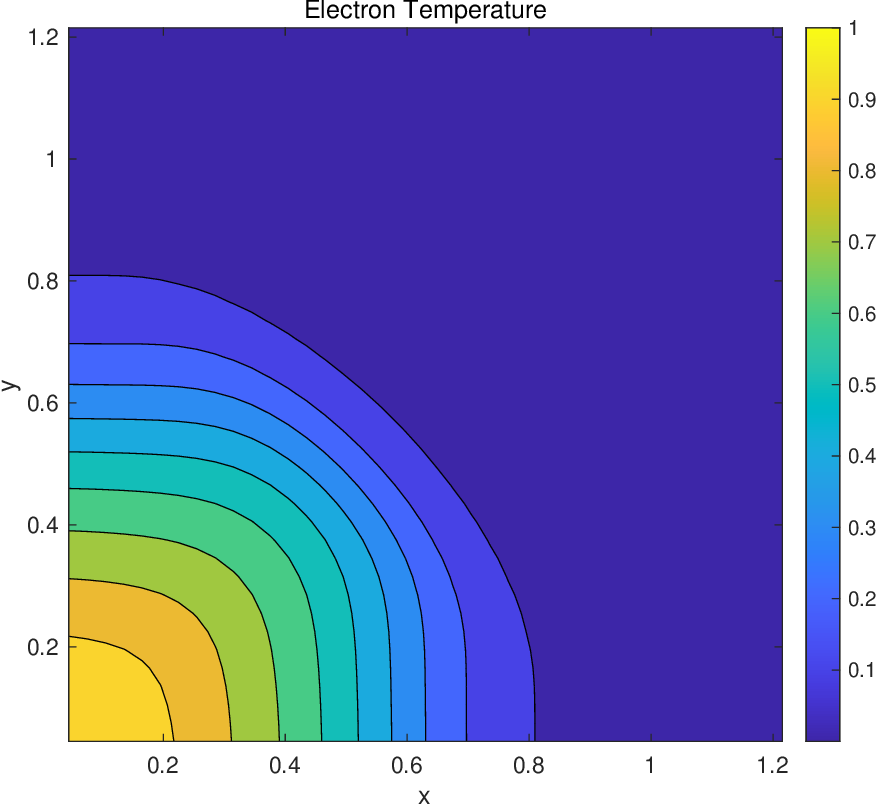}}	\subfloat[]{\includegraphics[width=0.33\textwidth]{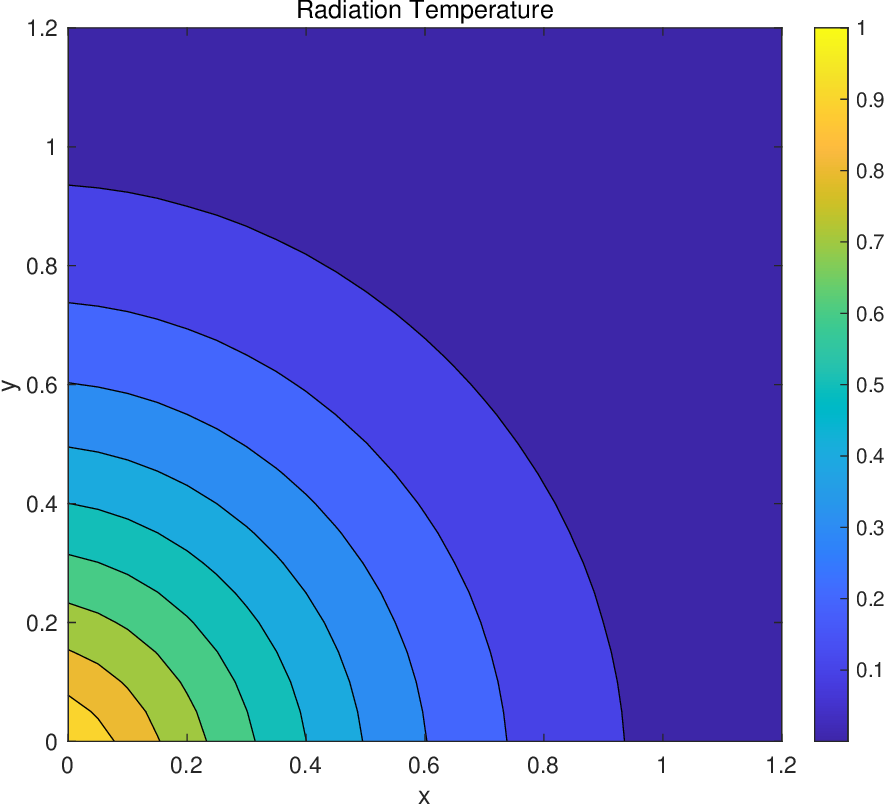}}
	\subfloat[]{\includegraphics[width=0.33\textwidth]{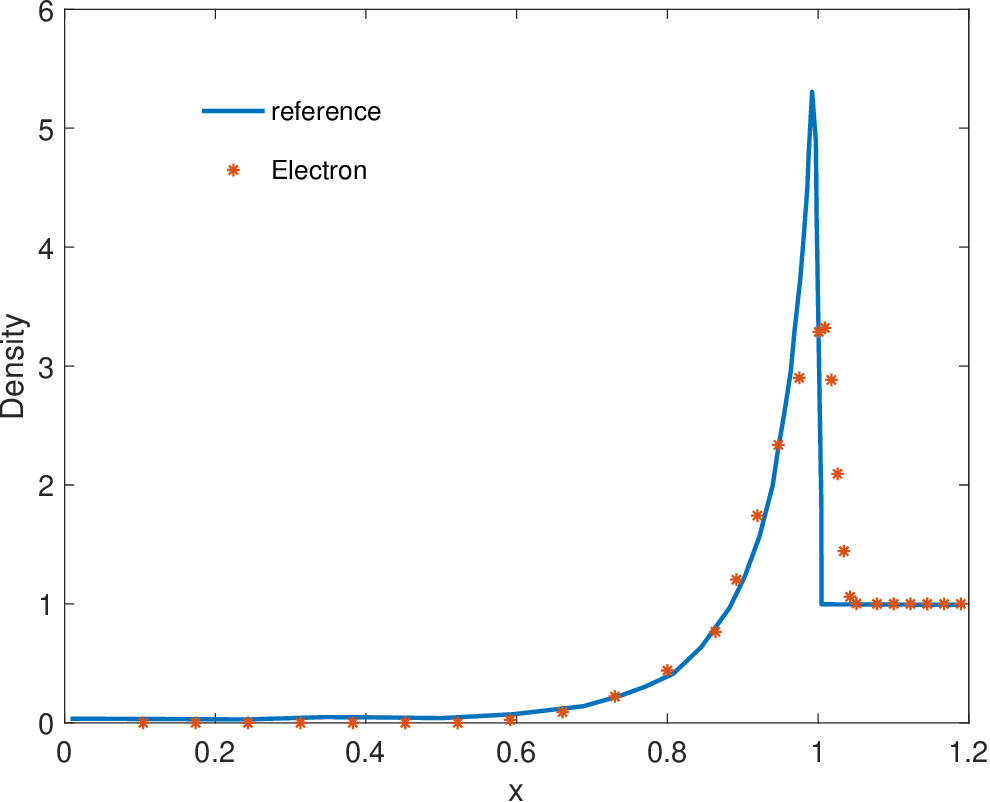}}
	\caption{Distributions of the (a) electron temperature, (b) radiaiton temperature, and (c) electron density for Sedov one point blast wave without exchange between electron and radiation.}
	\label{fig:sedov-non}
\end{figure}

\begin{figure}[H]
	\centering
	\subfloat[]{\includegraphics[width=0.33\textwidth]{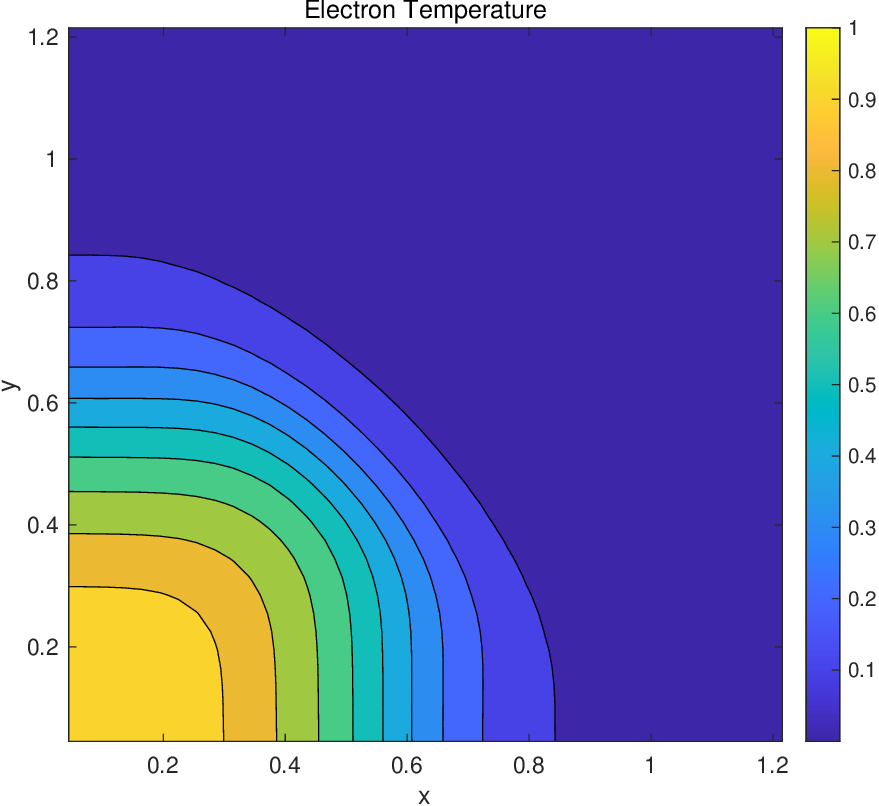}}	\subfloat[]{\includegraphics[width=0.33\textwidth]{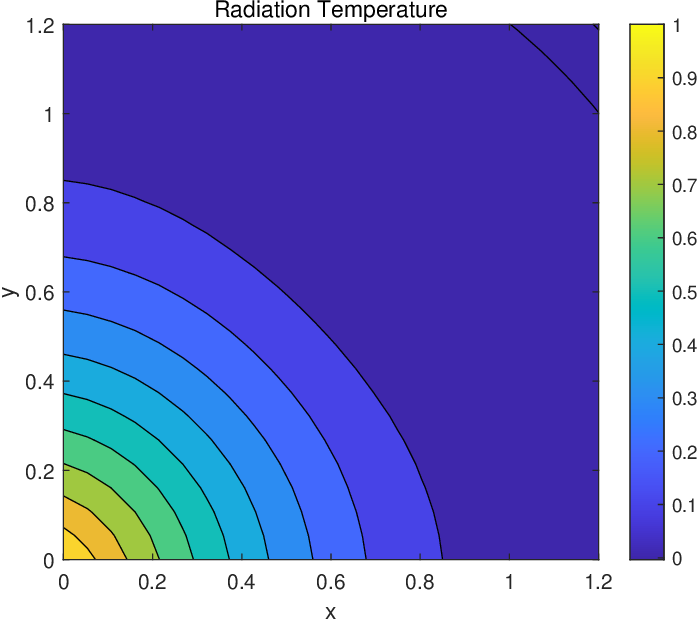}}
	\subfloat[]{\includegraphics[width=0.33\textwidth]{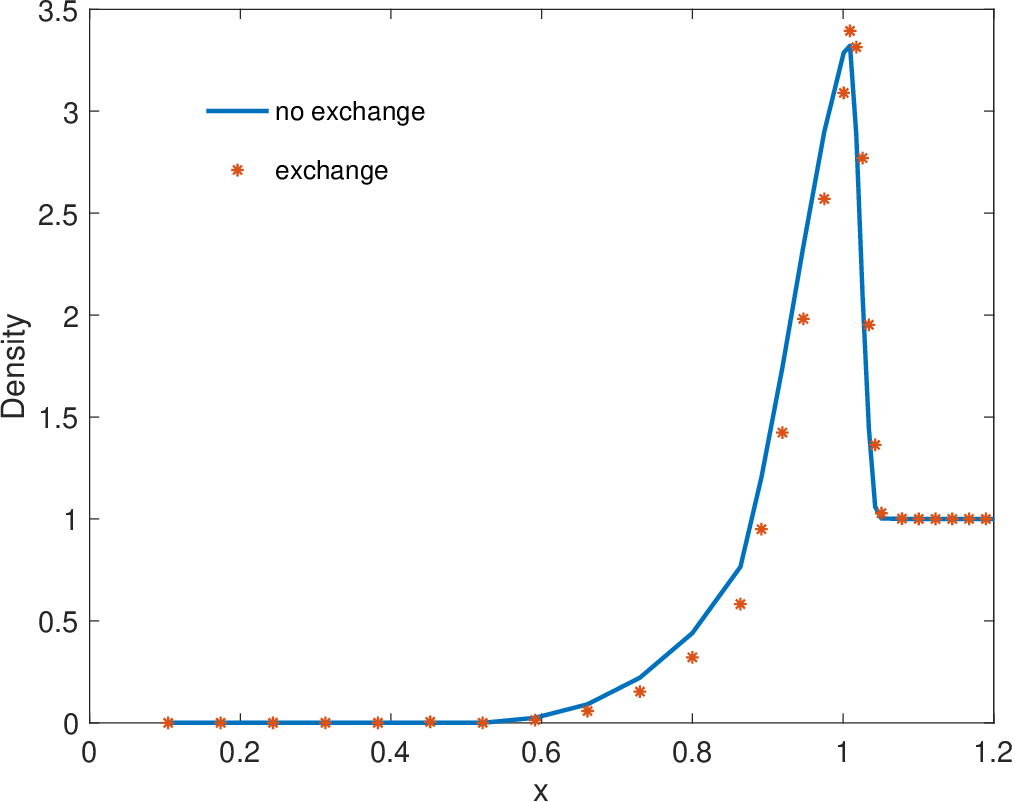}}
	\caption{Distributions of the (a) electron temperature, (b) radiaiton temperature, and (c) electron density for Sedov one point blast wave with exchange between electron and radiation.}
	\label{fig:sedov-ex}
\end{figure}

The 2-D Sedov problem is an explosion test case to model a blast wave from an energy-deposited singular point \cite{reinicke1991point}. The computational domain is [0,1.2] $\times$ [0,1.2]. Initial conditions with $\rho = 1, p = 10^{-4}, U = V = 0.0$  are imposed to the whole domain while $p = 0.106 * (\gamma - 1) / {\rm d}x {\rm d}y$ is set at [0,0]. The radiation intensity is initialized according to the electron temperature. And the velocity ratio $\vec{\beta} = \frac{\vec{U}}{c}$ is set as $0.01$, where the velocity of electron field $\vec{U}$ is selected as the maximum velocity when the density shock propagates to $r = 1.0$. The physical space is discretized with 80$\times$80 uniform meshes. The range of velocity space is $\Omega_x\in(-1,1), \Omega_y\in(0,2\pi)$ and the velocity space is with $30 \times 30$ meshes by midpoint rule. The electron density line is determined by connecting the origin to [1.2,1.2]. Fig.~\ref{fig:sedov-non} shows the Sedov case that does not involve exchanges between electron and radiation while Fig.~\ref{fig:sedov-ex} shows the result of the Sedov case with exchanges between electron and ion. The reference in Fig.~\ref{fig:sedov-non} is from \cite{kamm2007efficient}. By comparing the two cases, the influence of radiation on a one-point blast wave is analyzed. Compare Fig.~\ref{fig:sedov-non} and Fig.~\ref{fig:sedov-ex}, due to the interaction between the electron field and radiation field, electron temperature is propagating faster and the electron density shock front is also propagating faster.

\subsection{Tophat for radiation plasma system}
\begin{figure}[H]
	\centering
	\includegraphics[width=0.8\textwidth]{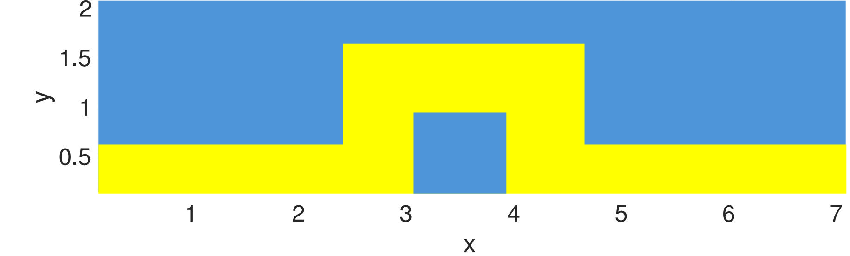}
	\caption{The demonstration of the initial settings of the two-dimensional Topha problem. Blue region represents the optical thick and continuum flow regime and yellow region represents the optical thin and free molecular flow regime. }
	\label{fig:demo-tophat}
\end{figure}

The two-dimensional Tophat problem is computed to verify the capability to capture the extreme non-equilibrium physics in both optically thick/thin regimes with different mass and number density ratios of ion and electron. In the Tophat case, strong discontinuity is set in terms of both ${\rm Kn}$ number and $\sigma$ number. Shown in Fig.~\ref{fig:demo-tophat}, the initial condition is set as an optically thick and continuum flow regime with $\sigma = 10^{2}, \rho = 1.0, p = 1.0,$ in $[x_1,x_2]\times[y_1,y_2] \in \{[3.0,4.0]\times[0,1.0], [0,2.5]\times[0.5,2.0], [4.5,7m_.0]\times[0.5,2.0], [2.5,4.5]\times[1.5,2.0]\}$, and an optical thin and free molecular flow regime with $\sigma = 10^{-1}, \rho = 10^{-5}, p = 10^{-5}$ otherwise. And the mass and number density ratio of electron and ion are $rM = \frac{m_{\mathcal{I}}}{m_\mathcal{E}}  = 4.0, \frac{n_{\mathcal{I}}}{n_\mathcal{E}+n_\mathcal{I}} = 0.2$.

The Knudsen number is defined by $L_{ref} = 1$, and the reference mean free path is calculated by the number density and molecular diameter for mean free path calculation is $n = n_{\mathcal{I}} + n_{\mathcal{E}}$ and $d = d_\mathcal{I} = d_\mathcal{E}$. The physical space $x \times  y = 7 \times 2$ is discretized by $70 \times 20$ uniform meshes. The velocity space is discretized by $30\times30$ meshes using the midpoint rule with the range $u_{\alpha} \in [-7,7]$ according to the most probable speed of each species $\alpha$.

\begin{figure}[H]
	\centering
	\centering
	\subfloat[]{\includegraphics[width=0.5\textwidth]{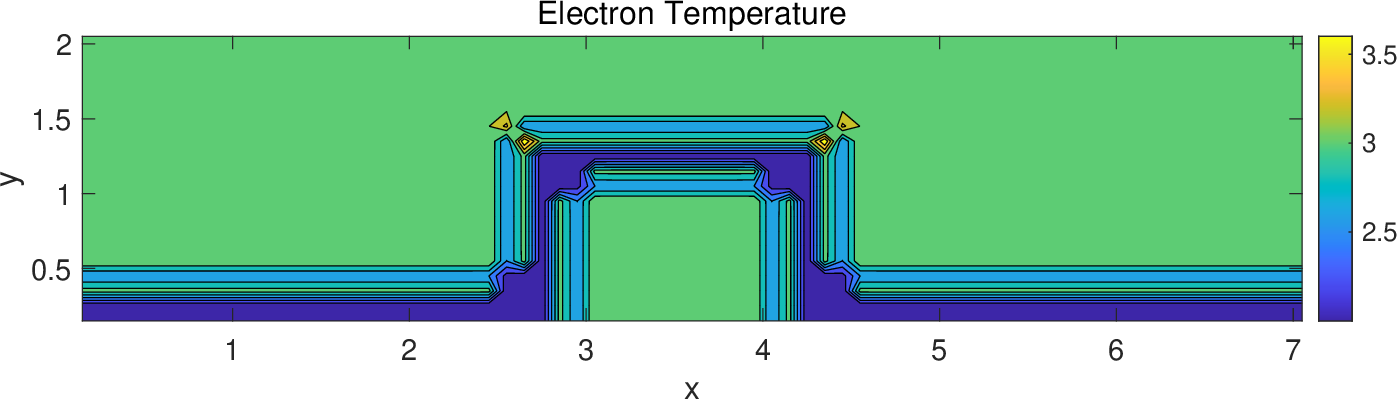}}	\subfloat[]{\includegraphics[width=0.5\textwidth]{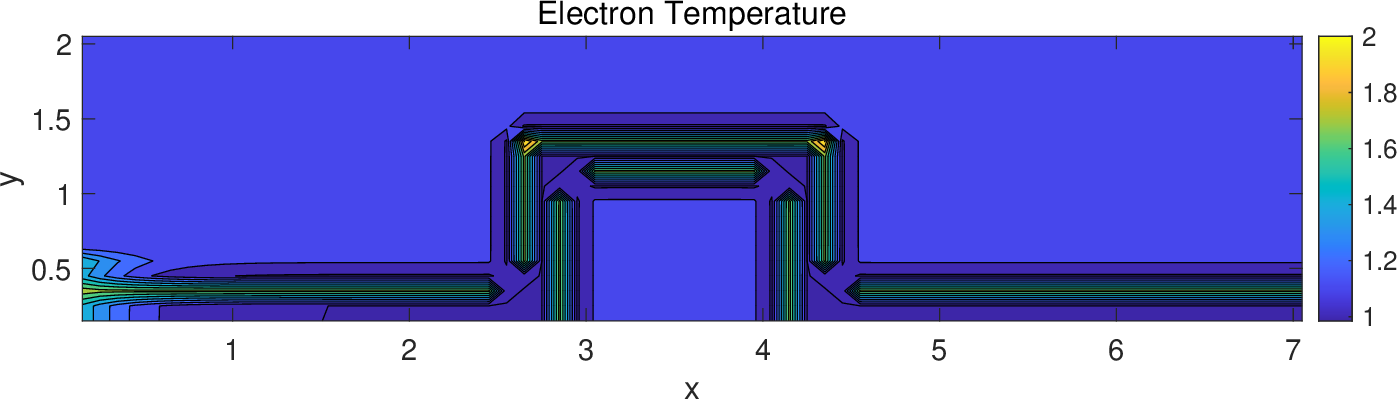}}

	\subfloat[]{\includegraphics[width=0.5\textwidth]{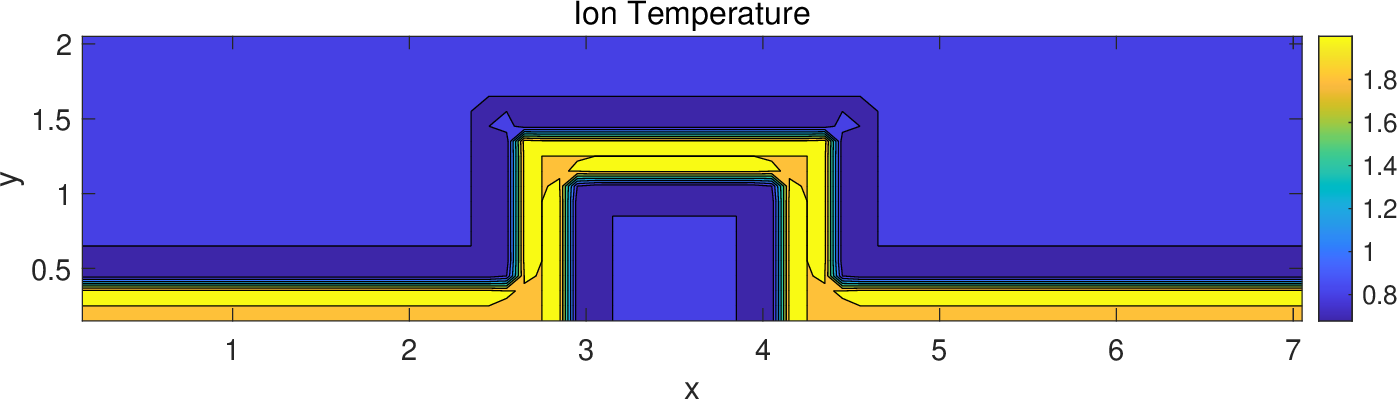}}	\subfloat[]{\includegraphics[width=0.5\textwidth]{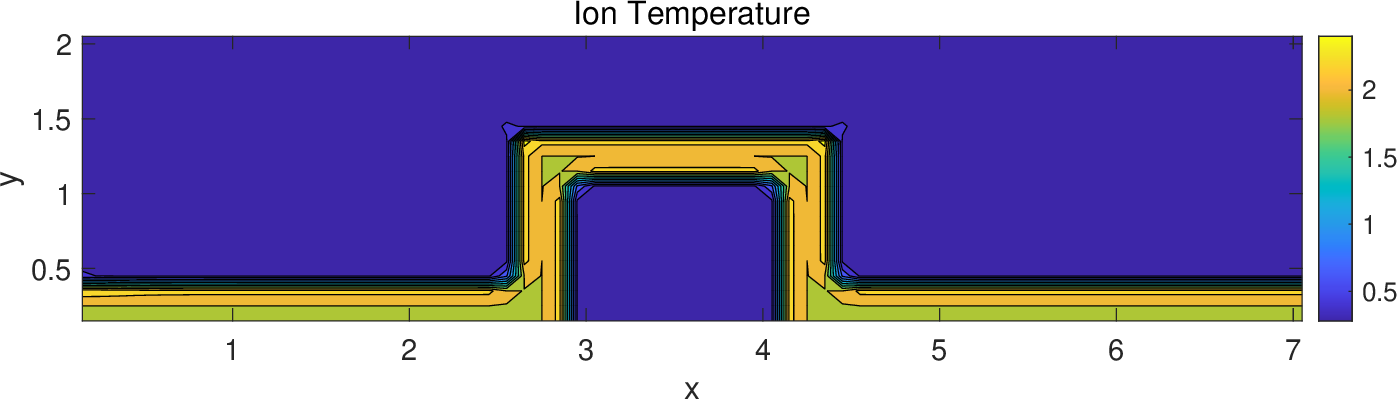}}
	
	\subfloat[]{\includegraphics[width=0.5\textwidth]{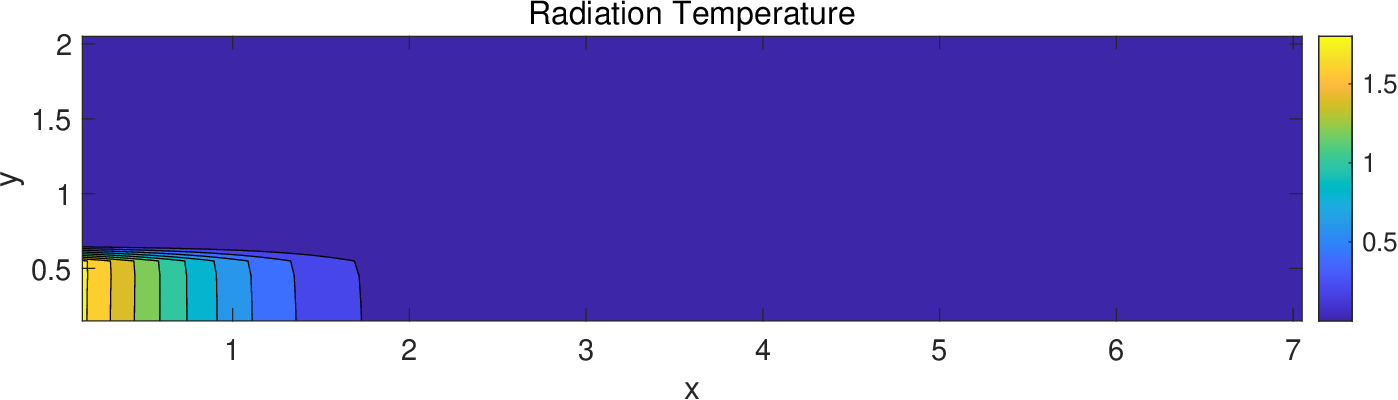}}
	\subfloat[]{\includegraphics[width=0.5\textwidth]{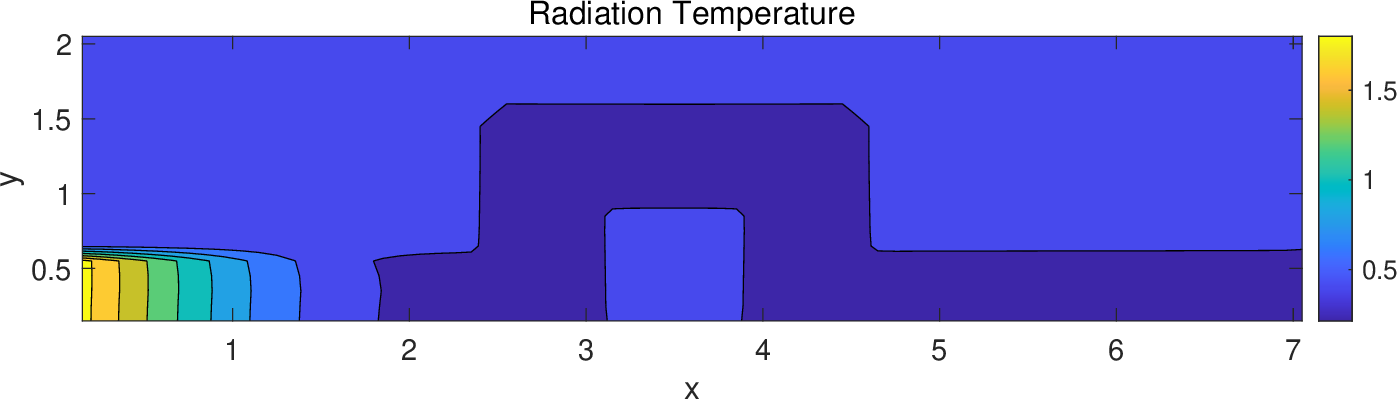}}
	\caption{Two-dimensional Tophat problem with the mass ratio $m_\mathcal{I}/m_\mathcal{E} = 4.0$ and number density ratio $n_\mathcal{I}/(n_\mathcal{E} + n_{\mathcal{I}} = 0.2$. Distributions of electron temperature, ion temperature, and  radiation temperature at $t = 0.01$. The left figure a,c,e are result without exchange with radiation field while the right figure b,d,f are result of radiation plasma sytem }
	\label{fig:tophat-0.01}
\end{figure}

\begin{figure}[H]
	\centering
	\centering
	\subfloat[]{\includegraphics[width=0.5\textwidth]{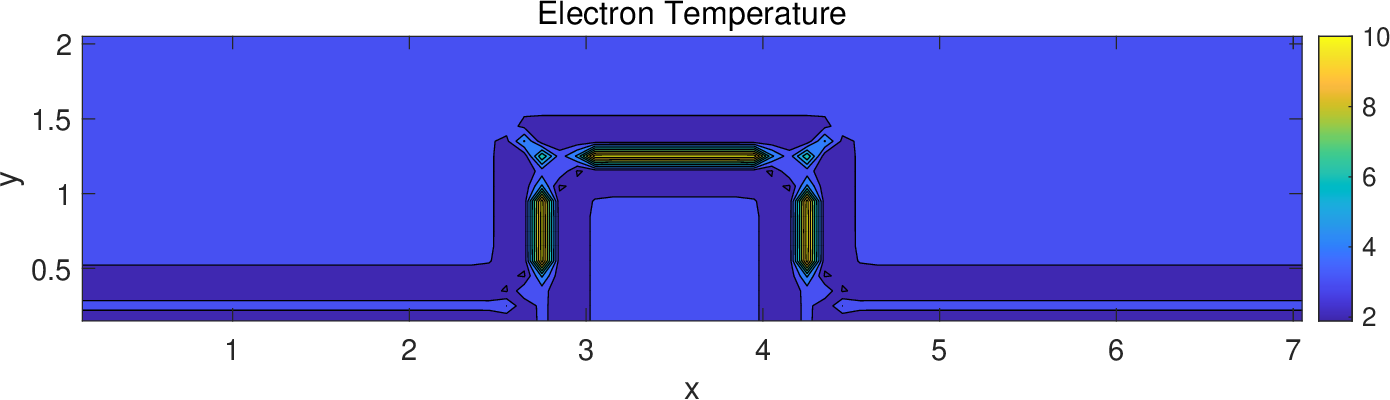}}	\subfloat[]{\includegraphics[width=0.5\textwidth]{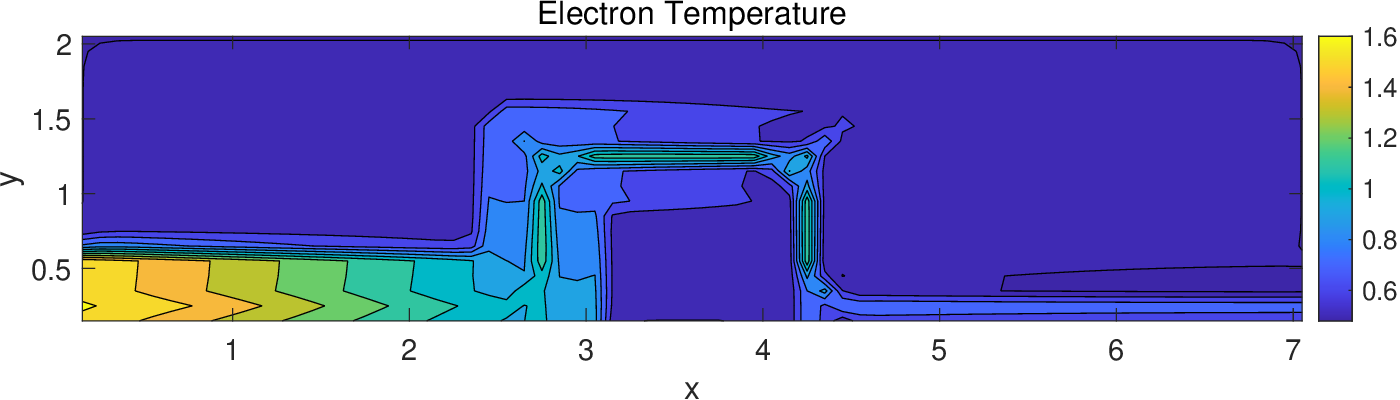}}
	
	\subfloat[]{\includegraphics[width=0.5\textwidth]{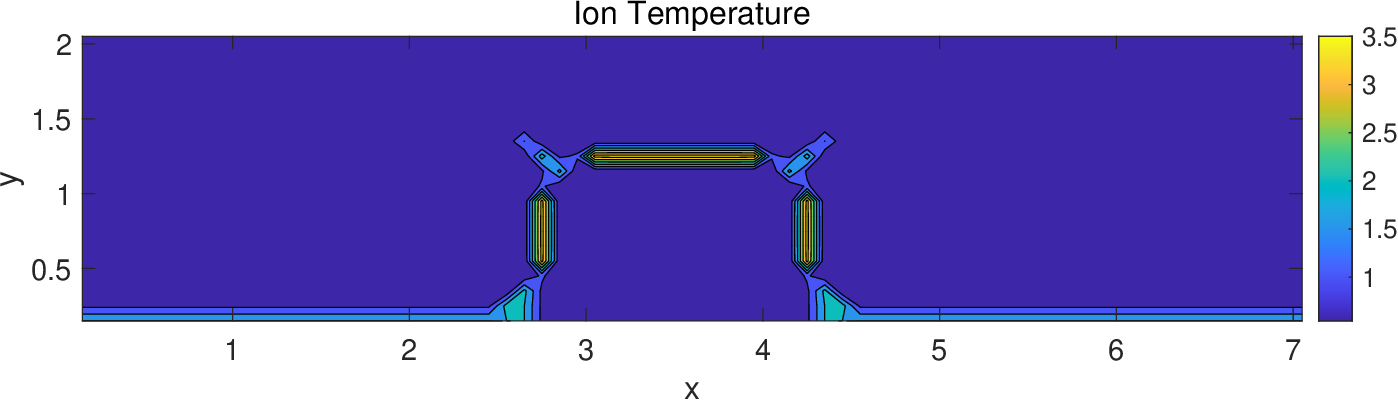}}	\subfloat[]{\includegraphics[width=0.5\textwidth]{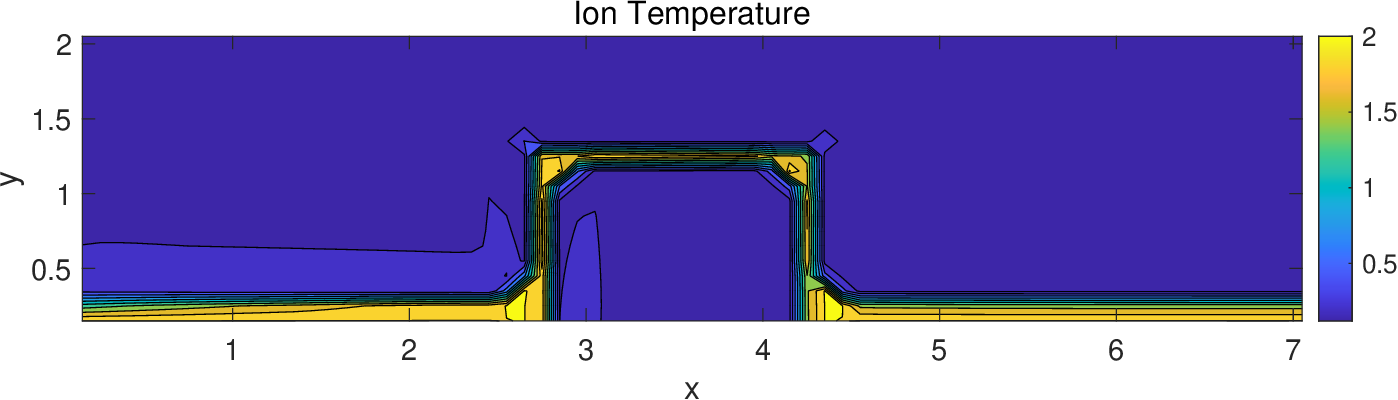}}
	
	\subfloat[]{\includegraphics[width=0.5\textwidth]{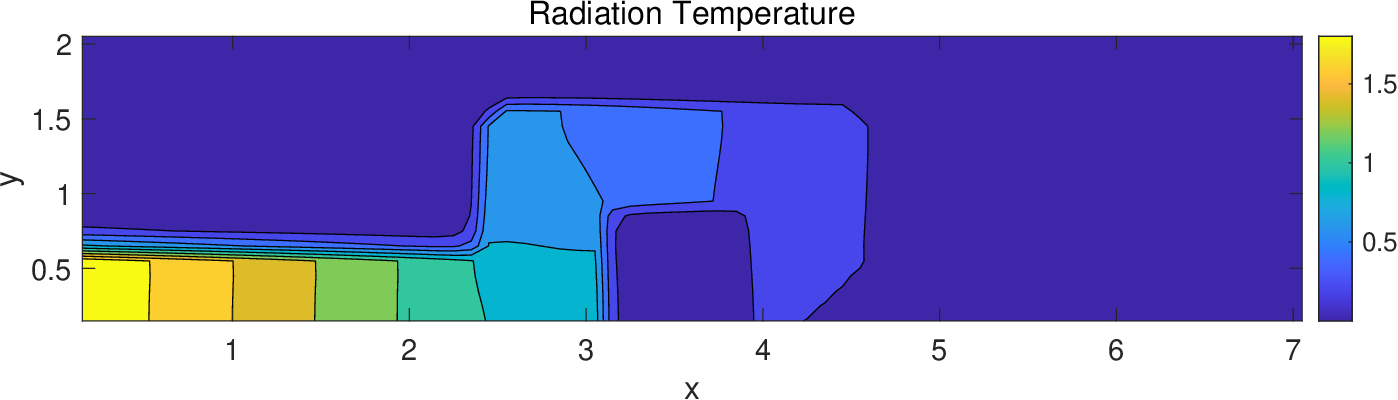}}
	\subfloat[]{\includegraphics[width=0.5\textwidth]{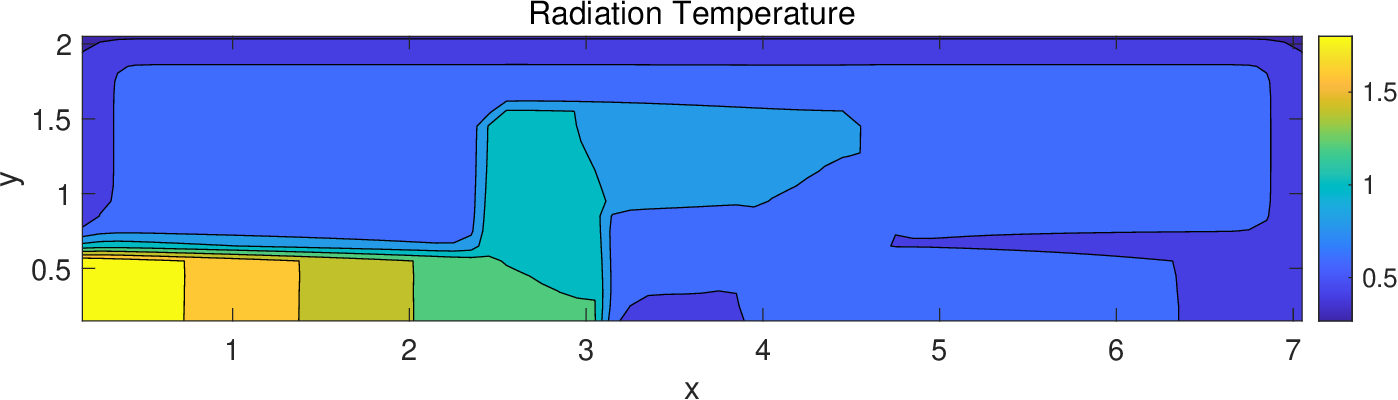}}
	\caption{Two-dimensional Tophat problem with the mass ratio $m_\mathcal{I}/m_\mathcal{E} = 4.0$ and number density ratio $n_\mathcal{I}/(n_\mathcal{E} + n_{\mathcal{I}} = 0.2$. Distributions of electron temperature, ion temperature, and  radiation temperature at $t = 0.1$. The left figure a,c,e are result without exchange with radiation field while the right figure b,d,f are result of radiation plasma sytem }
	\label{fig:tophat-0.1}
\end{figure}

\begin{figure}[H]
	\centering
	\centering
	\subfloat[]{\includegraphics[width=0.5\textwidth]{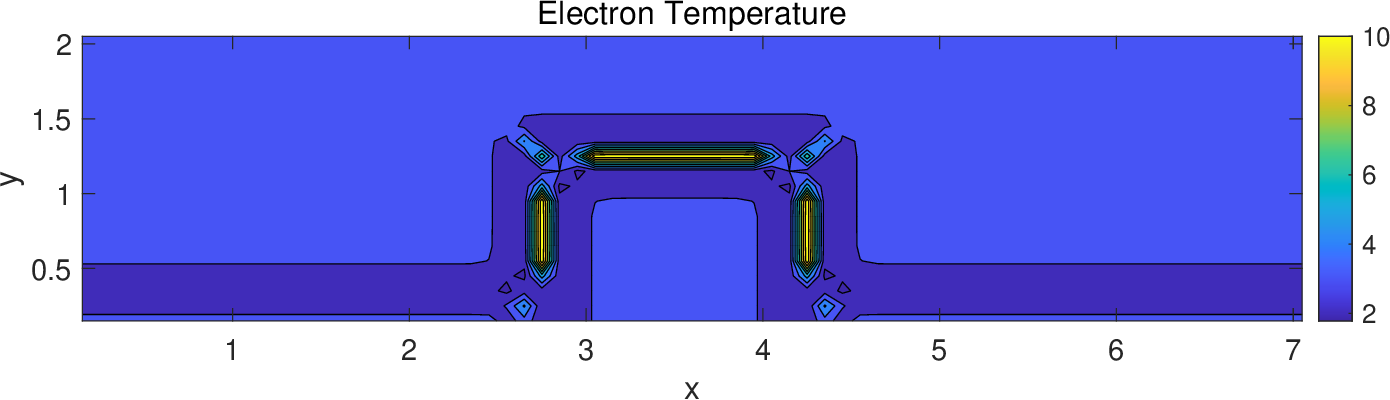}}	\subfloat[]{\includegraphics[width=0.5\textwidth]{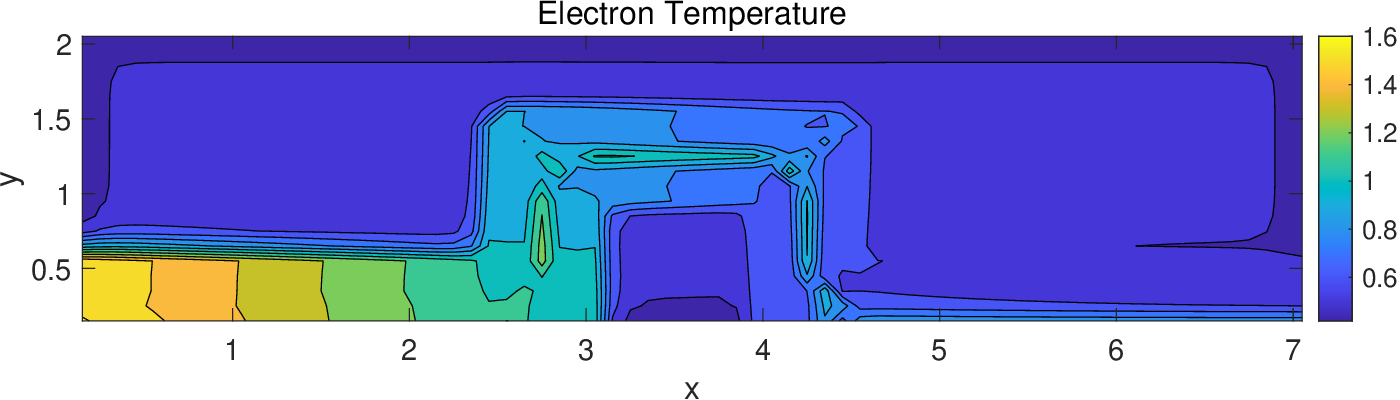}}
	
	\subfloat[]{\includegraphics[width=0.5\textwidth]{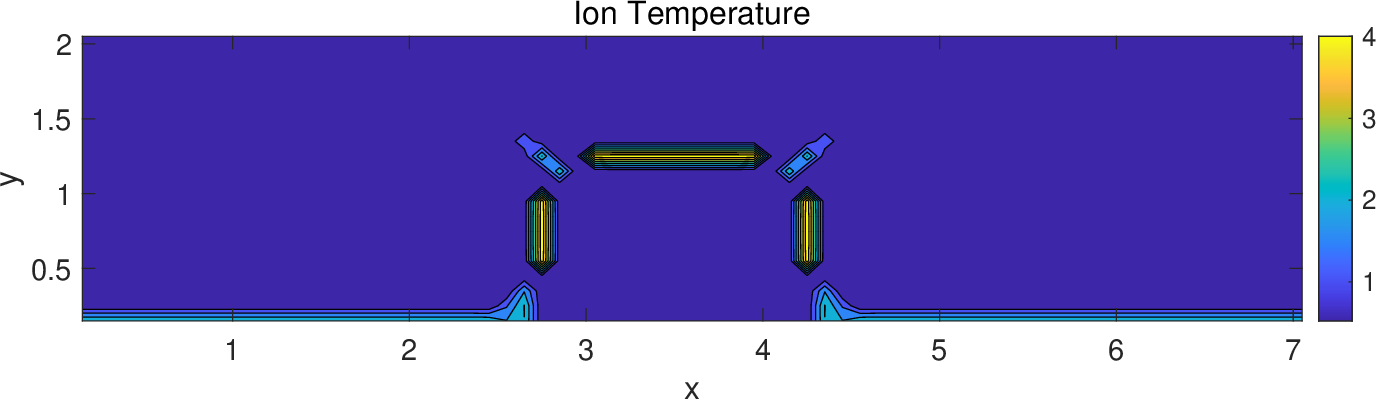}}	\subfloat[]{\includegraphics[width=0.5\textwidth]{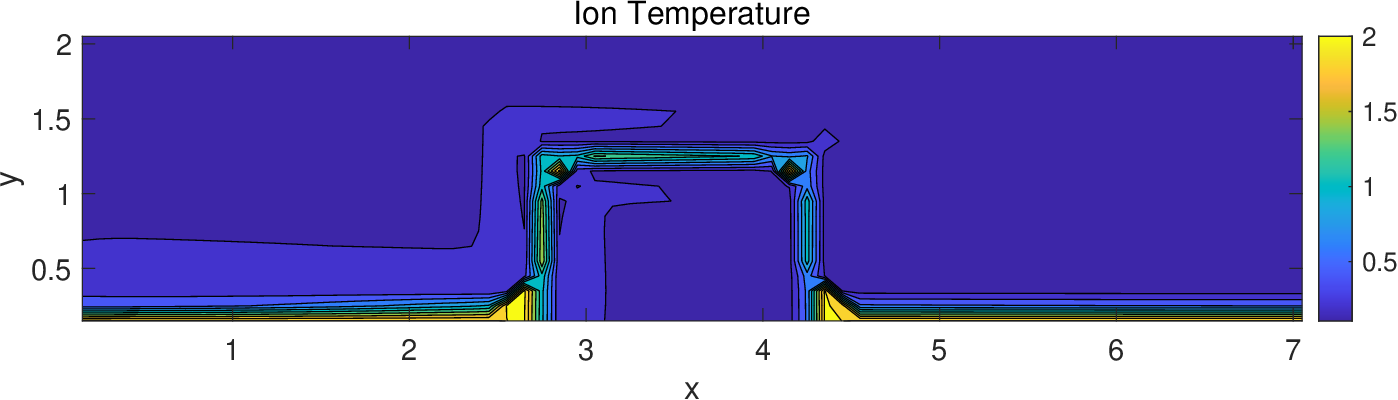}}
	
	\subfloat[]{\includegraphics[width=0.5\textwidth]{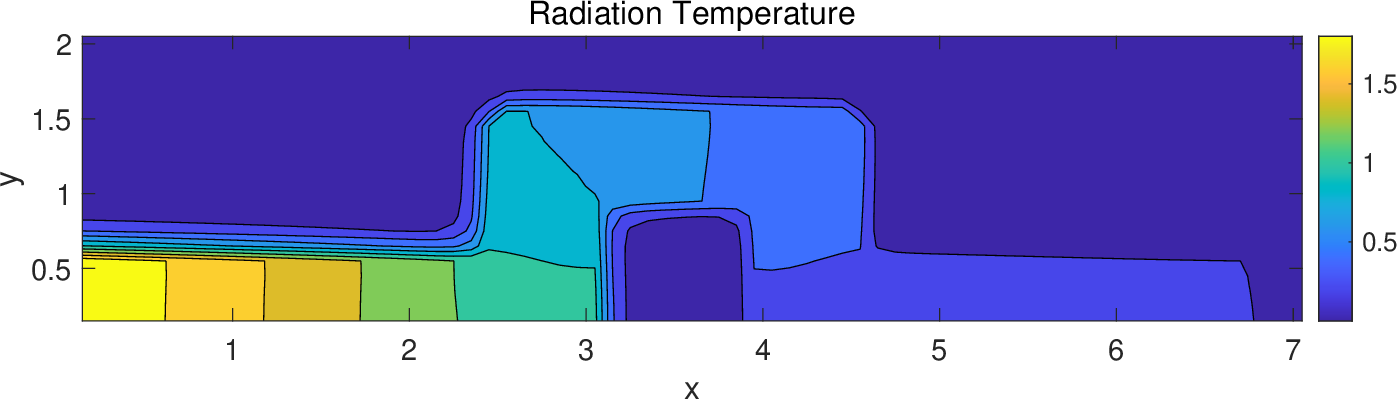}}
	\subfloat[]{\includegraphics[width=0.5\textwidth]{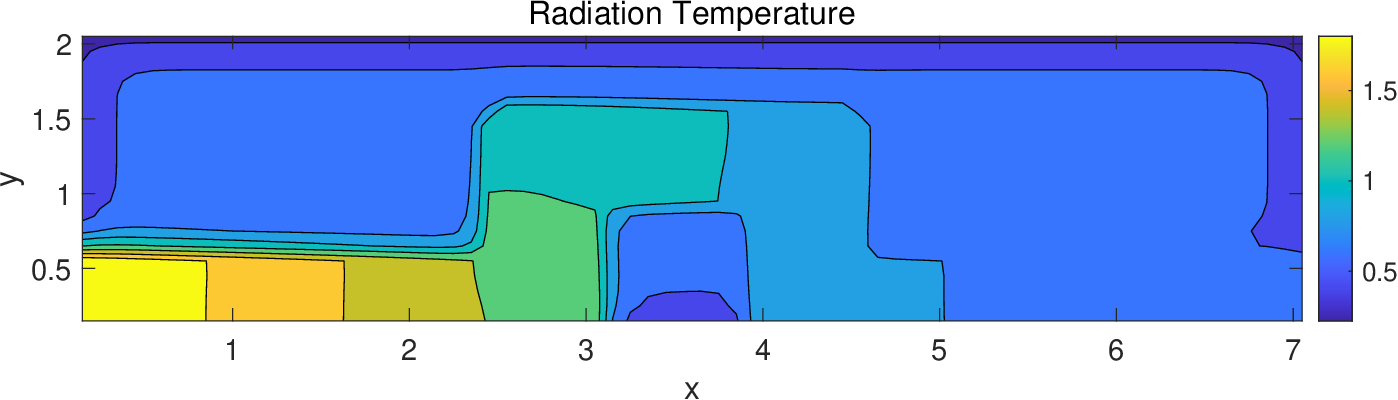}}
	\caption{Two-dimensional Tophat problem with the mass ratio $m_\mathcal{I}/m_\mathcal{E} = 4.0$ and number density ratio $n_\mathcal{I}/(n_\mathcal{E} + n_{\mathcal{I}} = 0.2$. Distributions of electron temperature, ion temperature, and  radiation temperature at $t = 0.2$. The left figure a,c,e are result without exchange with radiation field while the right figure b,d,f are result of radiation plasma sytem }
	\label{fig:tophat-0.2}
\end{figure}

\begin{figure}[H]
	\centering
	\centering
	\subfloat[]{\includegraphics[width=0.5\textwidth]{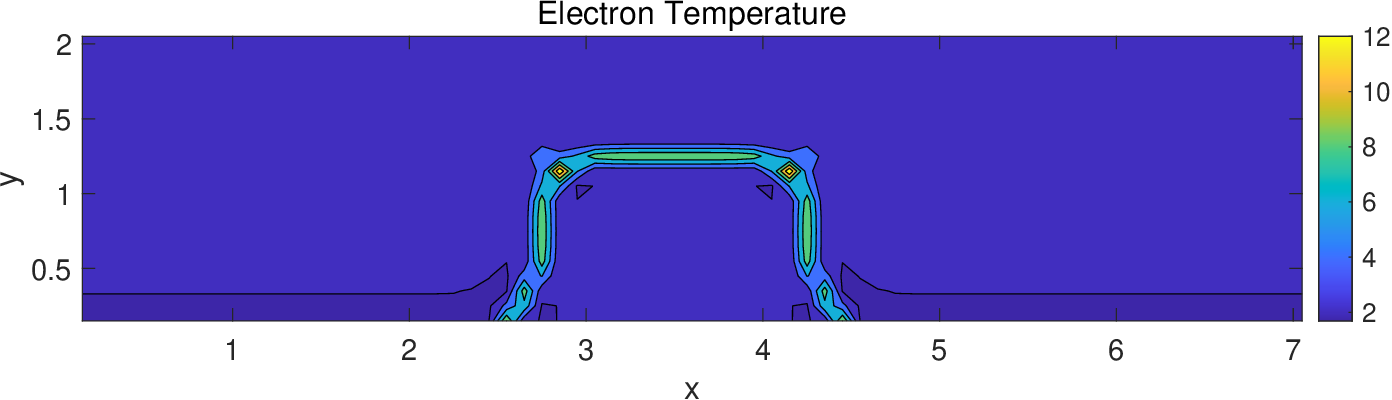}}	\subfloat[]{\includegraphics[width=0.5\textwidth]{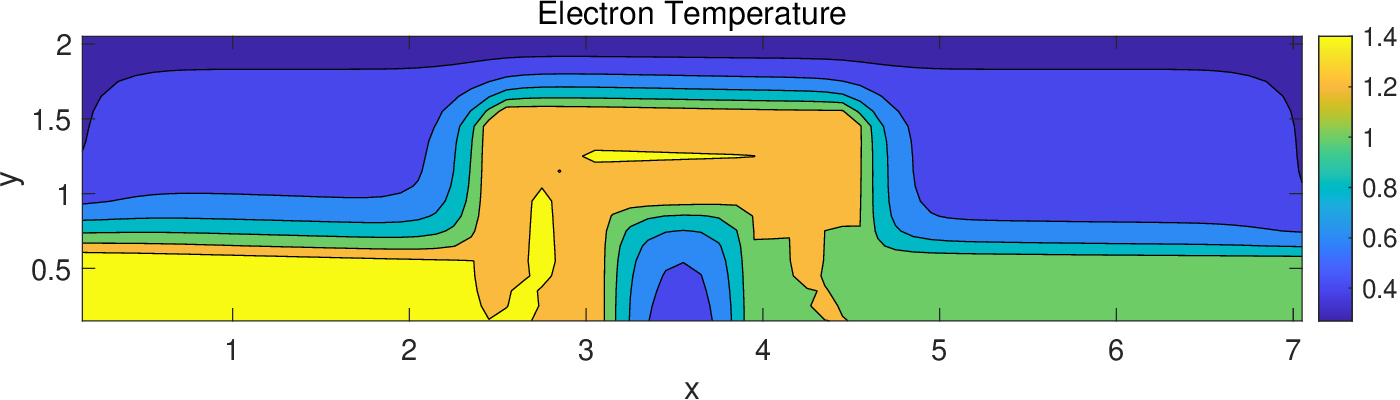}}
	
	\subfloat[]{\includegraphics[width=0.5\textwidth]{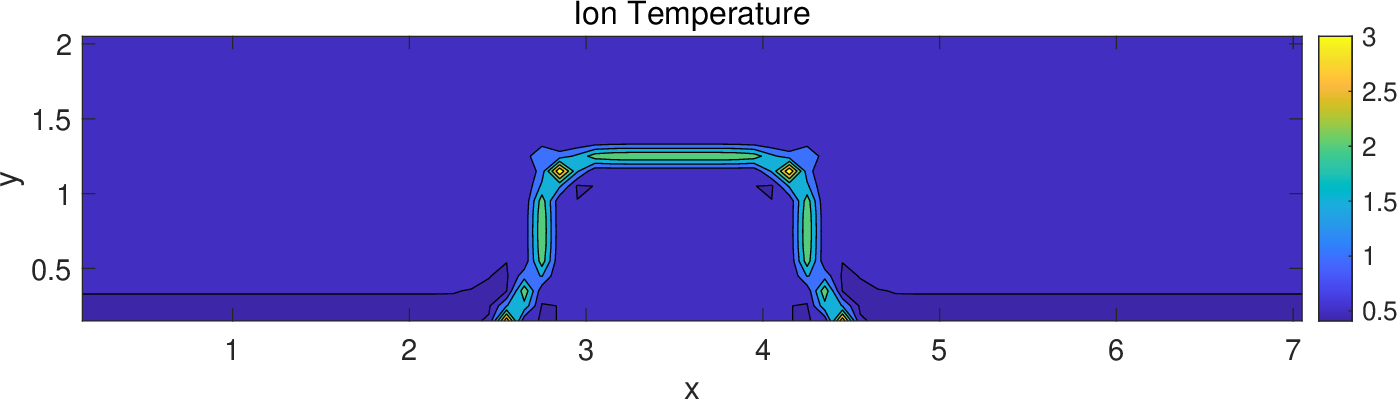}}	\subfloat[]{\includegraphics[width=0.5\textwidth]{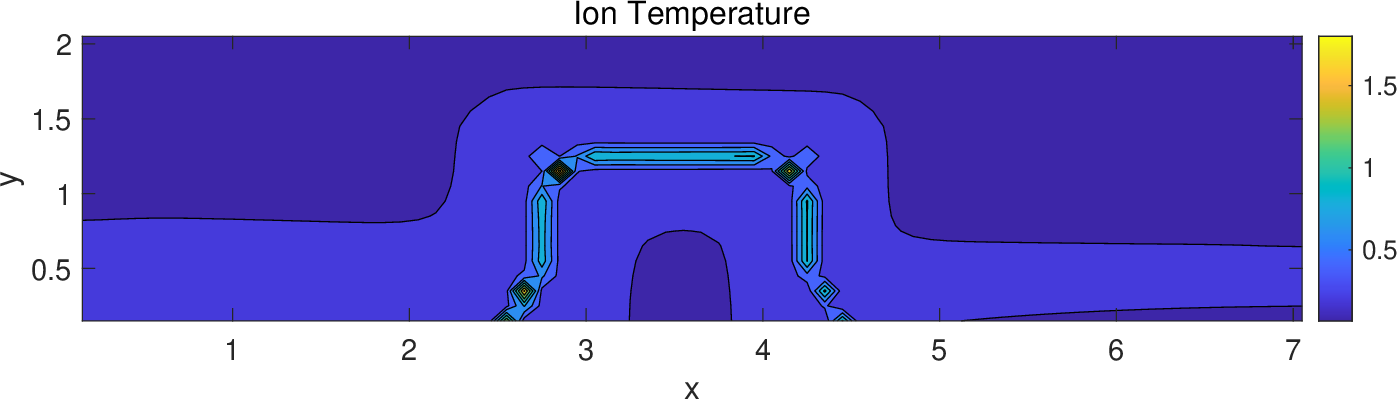}}
	
	\subfloat[]{\includegraphics[width=0.5\textwidth]{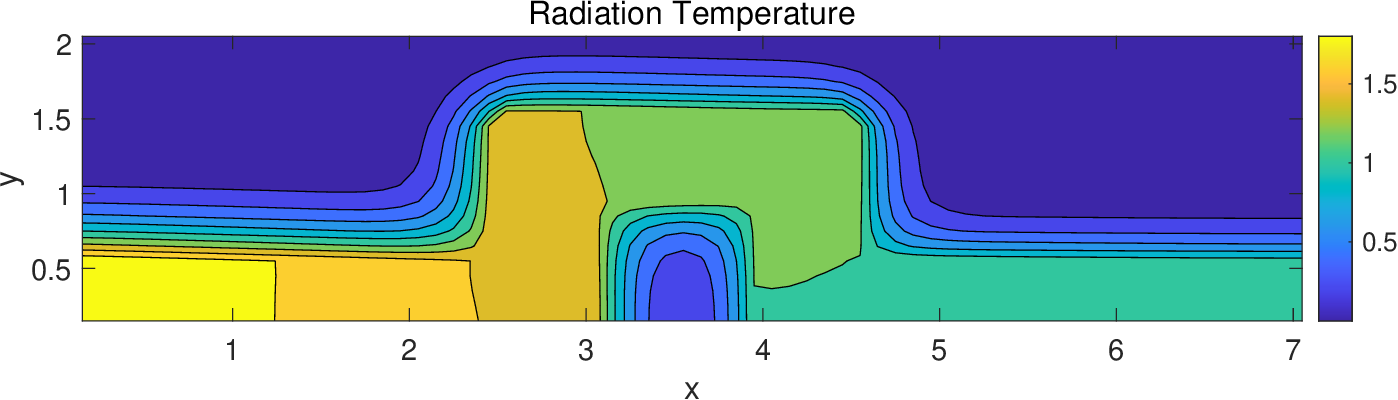}}
	\subfloat[]{\includegraphics[width=0.5\textwidth]{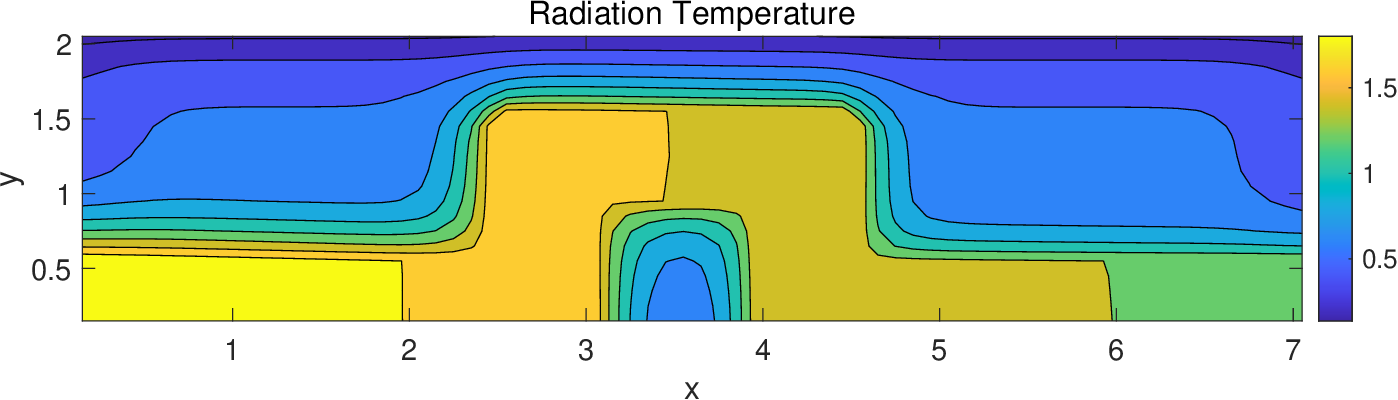}}
	\caption{Two-dimensional Tophat problem with the mass ratio $m_\mathcal{I}/m_\mathcal{E} = 4.0$ and number density ratio $n_\mathcal{I}/(n_\mathcal{E} + n_{\mathcal{I}} = 0.2$. Distributions of electron temperature, ion temperature, and  radiation temperature at $t = 1.0$. The left figure a,c,e are result without exchange with radiation field while the right figure b,d,f are result of radiation plasma sytem }
	\label{fig:tophat-1.0}
\end{figure}

\begin{figure}[H]
	\centering
	\subfloat[]{\includegraphics[width=0.33\textwidth]{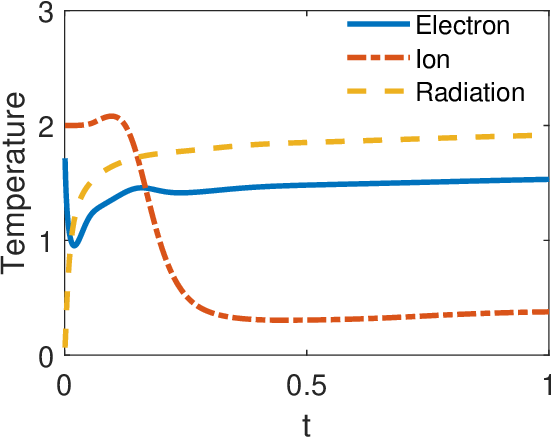}}
	\subfloat[]{\includegraphics[width=0.33\textwidth]{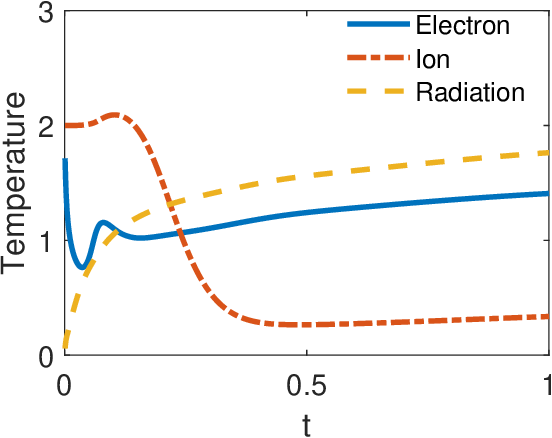}}
	\subfloat[]{\includegraphics[width=0.33\textwidth]{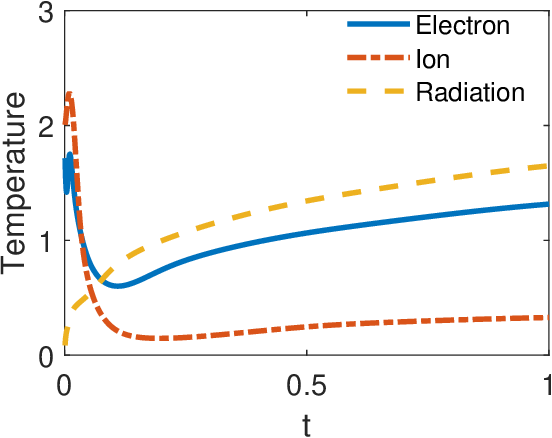}}
	\caption{Two dimensional Tophat problem with the mass ratio $m_\mathcal{I}/m_\mathcal{E} = 4.0$ and number density ratio $n_\mathcal{I}/(n_\mathcal{E} + n_{\mathcal{I}} = 0.2$ of ion to election. Time evolution of electron, ion and radiation temperature  from $ t = 0$ to $t = 0.25$ at $(x,y) = $  (a) $(1.0,0.25)$, (b) $ (2.75,0.25)$, and (c) $ (3.45,1.35)$.}
	\label{fig:tophat-piont}
\end{figure}
\begin{figure}[H]
	\centering
	\subfloat[]{\includegraphics[width=0.33\textwidth]{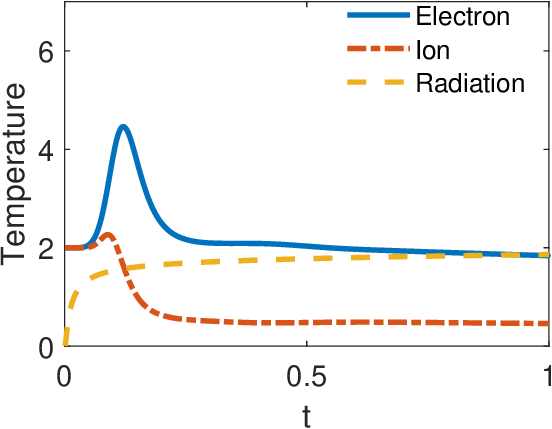}}
	\subfloat[]{\includegraphics[width=0.33\textwidth]{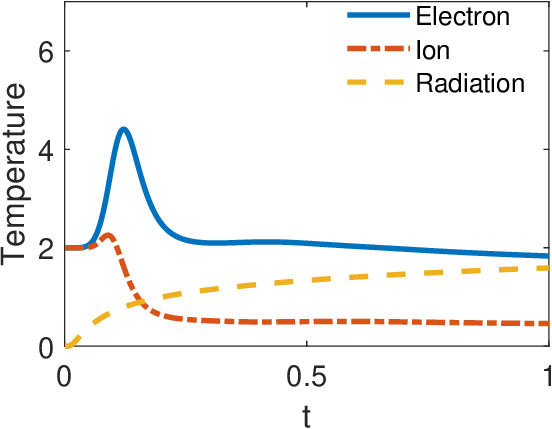}}
	\subfloat[]{\includegraphics[width=0.33\textwidth]{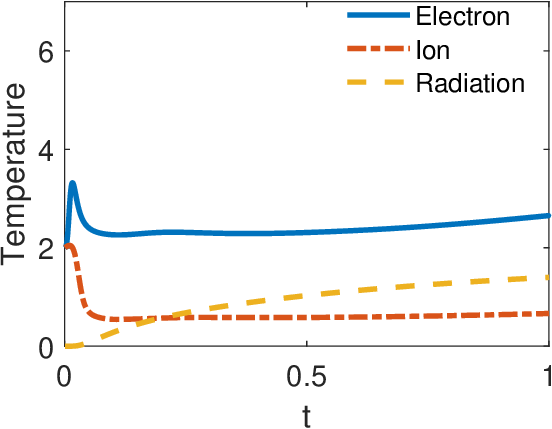}}
	\caption{Two dimensional Tophat problem with the mass ratio $m_\mathcal{I}/m_\mathcal{E} = 4.0$ and number density ratio $n_\mathcal{I}/(n_\mathcal{E} + n_{\mathcal{I}} = 0.2$ of ion to election without exchanges between electron and radiation. Time evolution of electron, ion and radiation temperature  from $ t = 0$ to $t = 0.25$ at $(x,y) = $  (a) $(1.0,0.25)$, (b) $ (2.75,0.25)$, and (c) $ (3.45,1.35)$.}
	\label{fig:tophat-piont1}
\end{figure}

The results at the time $t = 0.01, 0.1, 0.2, 1.0$ are chosen to illustrate the evolution of this problem. Firstly, the macroscopic density, velocity, and pressures of the ion and electron are the same. However, due to the differences in microscopic mass and number density, the temperature of the ion and electron deviate from each other, as shown in Fig.~\ref{fig:tophat-0.01}. Additionally, it can be seen that at an early stage of the evolution, the strong discontinuity across large scales in fluid field evolve quickly through the interaction wave at four corners for both ion and electron. From the comparison of Fig.~\ref{fig:tophat-0.1}, it is clear that the evolution of the radiation field first affects the electron temperature, and then the ion temperature is affected by the electron temperature. When the perturbation of radiation reaches to the corner where the interaction of electron and ion is intensive, the symmetric structure of the corners is broken, as shown in Fig.~\ref{fig:tophat-0.2}. Also the result shows that when the radiation temperature strongly influences the electron and ion temperatures, the interaction waves induced by strong discontinuity are dissipated during the evolution. The influence of radiation temperature dominates the flow fields. In this case, the capability of our model to resolve the non-equilibrium flows which includes the multiscale transfer process is validated.  The Fig.~\ref{fig:tophat-1.0} shows that, as time evolves, the radiation temperature undergoes a gradual evolution towards stability, accompanied by a progressive restoration of a symmetrical structure in the electron and ion temperatures.
Fig.~\ref{fig:tophat-piont} and Fig.~\ref{fig:tophat-piont1} shows the time evolution of electron, ion and radiation temperature at three chosen point, $(x,y)= (1.0, 0.25), (2.75, 0.25),  (3.45,1.35)$, to give a more clear picture of interaction among electron, ion and radiation in the strong non-equilibrium flow region.  It is clearly that due to the energy exchange, the pattern of electron and ion temperature evolution is deviated at three different points. Also,  after the radiation temperature propagates to that point, the electron temperature increases first, followed by driving up the ion temperature.

\section{Conclusion}\label{sec:conclusion}
In this paper, we present an extended gas kinetic-unified gas kinetic scheme for modeling the time-evolution of coupled radiation, electron, and ion energies in multiscale radiation plasma transport physics.
This system treats electrons and ions as a dual fluid model and radiation as a nonequilibrium transport model with the exchange of momentum and energy with electrons and ions. 
The hydrodynamics limit of the method is consistent with the traditional radiation hydrodynamic equations. Furthermore, due to the
multiscale transport for the radiation, the current method can capture the radiative transfer in the entire regime from optically thick to optically thin. 
Overall, the proposed scheme resolves radiation plasma system with exhibition of multiscale properties. 
Many test cases have been used to validate the scheme. 
This method offers a computational tool for simulating radiation plasma problem in astrophysics and inertial confinement fusion across various energy density regimes.

\section*{Author's contributions}

All authors contributed equally to this work.

\section*{Acknowledgments}

This current work was supported by National Key R$\&$D Program of China (Grant Nos. 2022YFA1004500),
National Natural Science Foundation of China (12172316, 92371107),
and Hong Kong research grant council (16208021,16301222).

\section*{Data Availability}

The data that support the findings of this study are available from the corresponding author upon reasonable request.




\bibliographystyle{elsarticle-num}
\bibliography{ref.bib}







\end{document}